%% file: ms.tex
\documentclass[prodmode,acmjetc]{acmsmall} 
\pdfoutput=1
\usepackage{makecell} 				
\usepackage{subfigure}
\usepackage{graphicx}
\usepackage{fixltx2e}
\usepackage{subfig}
\usepackage{csquotes}
\usepackage{cite}
\usepackage{caption}
\usepackage{subfig}
\usepackage{amssymb}
\usepackage{amsmath}
\usepackage{siunitx}
\usepackage{microtype}
\usepackage{upgreek}

\usepackage{xcolor,colortbl}
\definecolor{Gray}{gray}{0.95}
\newcolumntype{a}{>{\columncolor{Gray}}c}
\usepackage{xcolor,soul}
\definecolor{light-gray}{gray}{0.90}
\sethlcolor{light-gray}

\usepackage{enumitem,booktabs}
\usepackage[referable]{threeparttablex}
\renewlist{tablenotes}{enumerate}{1}
\makeatletter
\setlist[tablenotes]{label=\tnote{\alph*},ref=\alph*,itemsep=\z@,topsep=\z@skip,partopsep=\z@skip,parsep=\z@,itemindent=\z@,labelsep=.2em,leftmargin=*,align=left,before={\footnotesize}}
\makeatother

\graphicspath{{Figures/}}
\newcommand{\eq}[1]{Equation~(\ref{#1})}
\newcommand{\fig}[1]{Figure~\ref{#1}}
\newcommand{\tb}[1]{Table~\ref{#1}}

\usepackage[ruled]{algorithm2e}

\SetAlFnt{\small}
\SetAlCapFnt{\small}
\SetAlCapNameFnt{\small}
\SetAlCapHSkip{0pt}
\IncMargin{-\parindent}

\acmVolume{1}
\acmNumber{1}
\acmArticle{1}
\acmYear{2018}
\acmMonth{1}

\doi{10.1145/3233987}

\issn{000-0000}

\begin{document}

\markboth{A. Zyarah et al.}{Semi-Trained Memristive Crossbar Computing Engine with In-Situ Learning Accelerator}

\title{Semi-Trained Memristive Crossbar Computing Engine with In-Situ Learning Accelerator}
\author{Abdullah M. Zyarah
\affil{Neuromorphic AI Lab, Rochester Institute of Technology}
Dhireesha Kudithipudi
\affil{Neuromorphic AI Lab, Rochester Institute of Technology \vspace{-2mm}}}

\begin{abstract}
\hrule \vspace{2mm}
On-device intelligence is gaining significant attention recently as it offers local data processing and low power consumption. In this research, an on-device training circuitry for threshold-current memristors integrated in a crossbar structure is proposed. Furthermore, alternate approaches of mapping the synaptic weights into fully-trained and semi-trained crossbars are investigated. In a semi-trained crossbar a confined subset of memristors are tuned and the remaining subset of memristors are not programmed. This translates to optimal resource utilization and power consumption, compared to a fully programmed crossbar. The semi-trained crossbar architecture is applicable to a broad class of neural networks. System level verification is performed with an extreme learning machine for binomial and multinomial classification. The total power for a single 4x4 layer network, when implemented in IBM 65nm node, is estimated to be $\approx$ 42.16$\upmu$W and the area is estimated to be 26.48$\upmu$m x 22.35$\upmu$m.
\end{abstract}


\setcopyright{acmcopyright}
\acmYear{2018} \acmVolume{x} \acmNumber{x} \acmArticle{x} \acmMonth{1} 

\terms{Design, Analysis, Experimentation}

\acmformat{Abdullah M. Zyarah and Dhireesha Kudithipudi, 2018. Semi-Trained Memristive Crossbar Computing Engine with In-Situ Learning Accelerator.}

\begin{bottomstuff}
This material is based on research sponsored by AirForce Research Laboratory under agreement number FA8750-16-1-0108. The U.S. Government is authorized to reproduce and distribute reprints for Governmental purposes notwithstanding any copyright notation thereon. The views and conclusions contained herein are those of the authors and should not be interpreted as necessarily representing the official policies or endorsements, either expressed or implied, of AirForce Research Laboratory or the U.S. Government.

Authors’ addresses: A. M. Zyarah and D. Kudithipudi, Neuromorphic AI Lab, Rochester Institute of Technology, Rochester, NY; emails: \{amz6011, dxkeec\}@rit.edu.
\end{bottomstuff}
\maketitle
\section{Introduction}
On-device intelligence is gaining significant attention recently as it offers local data processing and low power consumption, suitable for energy constrained platforms (\textit{e.g. } IoT). Porting neural networks on to embedded platforms to enable on-device intelligence requires high computational power and bandwidth. Conventional architectures, such as von Neumann architecture, suffer from throughput drop and high power draw when realizing neural networks. This can be attributed to the physical separation between processing and memory units, which leads to memory bottleneck~\cite{indiveri2015memory}. Additionally, pure CMOS implementation of neural networks impose area and power constraints that hinder the deployment on to embedded platforms~\cite{kim2012neural}. 

In 2008, a successful physical implementation of a synapse-like device called memristor was proposed by Strukov~\cite{strukov2008missing}. Theoretically, the memristor was introduced by L. Chua in 1971 as a fourth fundamental electrical device that correlates the flux and charge in a non-linear relationship~\cite{chua1971memristor}. The memristor acts like a non-volatile memory element~\cite{borghetti2010memristive}, consumes low  energy~\cite{prezioso2015training,merkel2017current}, has a small footprint compared to transistors, and can be integrated in high density crossbar structures~\cite{jo2010nanoscale}. A key advantage of the crossbar structure is that it enables performing the most computationally intensive operations (multiply-accumulate) in neural networks concurrently while consuming small amount of power compared to conventional implementations~\cite{snider2008spike,taha2014memristor}. These properties make the memristor a natural choice for realizing neural networks in an efficient manner such that it meets embedded device constraints. Typically, memristive devices are used to model the bi-polar synaptic weights in neural networks. Due to the fact that a memristor exhibits properties similar to that of a resistor, memristors can represent only positive range of weights. Thus, either a hybrid CMOS-memristor~\cite{kim2012neural,soudry2015memristor} or two memristors~\cite{alibart2013pattern,hu2014memristor} are used to model the bipolar synaptic weights. Although modeling the synaptic weight with one memristor is easier to train, it demands additional circuitry to generate the bipolar weights. On the other hand, using two memristors to model the synaptic weights reduces the power consumption, but increases the hardware complexity and the training process.

Several research groups have studied the realization of synaptic weights in memristive devices while enabling the on-device learning. To realize the synaptic weight with one memristor, Sah et al. proposed a memristor-based synaptic circuit which employs an H-Bridge and doublet generator to perform positive and negative input-weight multiplication. However, it was not studied in the context of a multi-level network or a crossbar architecture~\cite{sah2012memristor}. In 2015, Soudry et al. presented a memristor crossbar that supports on-chip online gradient descent. In this architecture, two transistors and a memristor were used to implement a synapse. This makes the total number of transistors in the crossbar scale linearly with the number of memristors~\cite{soudry2015memristor}. Adopting two memristors to model the synaptic weights is studied by Alibart et al. who proposed a memristor-based single-layer perceptron to classify synthetic pattern of the letters 'X' and 'T'. The proposed design is trained using ex-situ and in-situ methods~\cite{alibart2013pattern}. Hasan et al. presented an on-chip training circuit to account for device faults and variability in memristor-based deep neural networks~\cite{hasan2017chip}. This network was trained with auto-encoders and backpropagation and simulated in MATLAB for classification application. When it comes to extreme learning machine (ELM), which is the neural network algorithm used to verify our proposed architecture, few research groups have studied the memristor-based ELM. In 2014 Merkel et al. proposed memristor-based ELM implementation, but it is not studied within the context of a crossbar structure~\cite{merkel2014neuromemristive}. Later in 2015, OxRAM based ELM architecture was proposed by Suri et al. in which the nano-scale device variability is exploited to design ELM in an efficient manner ~\cite{suri2015oxram}. Unfortunately, this work does not provide details about the hardware implementation and the training process. It is also important to mention here that most memristor-based neural network architectures proposed in literature use threshold-voltage memristors. To the best of our knowledge, no design has explored on-device learning for current-threshold memristors integrated in crossbar architecture.

This paper proposes on-device training circuitry for current threshold memristors integrated into a crossbar structure. Moreover, the paper presents a different approach for realizing the synaptic weights into a memristive crossbar such that bipolar weights are obtained. The proposed approach is based on using semi-trained crossbar structure (a combination of trained and untrained memristors), where the trained memristor models the synaptic weights and the fixed ones are used in association with the trained memristor to generate bipolar synaptic weights. The proposed design is simulated in Cadence Spectre and verified for classification application in MATLAB using binomial (Diabetes and Australian Credit) and multinomial (Iris and MNIST) datasets ~\cite{Lichman:2013,lecun1998mnist}. For a single 4x4 layer network (crossbar and its associated control and training circuitries) implemented in IBM 65nm technology node, the total power is estimated to be $\approx$ 42.16$\upmu$W, while the area is 26.48$\upmu$m x 22.35$\upmu$m.

The rest of the paper is organized as follows: Section 2 presents an overview about ELM. Section 3 and 4 discuss the design methodology and the hardware analysis. The experimental setup is described in Section 5. Section 6 demonstrates the experimental results and Section 7 concludes the paper.

\section{Overview of ELM}
Extreme learning machine (ELM) is a multi-layer feed-forward neural network used in real-time regression and classification applications~\cite{huang2004extreme}. It has roots back in the random vector functional link (RVFL) networks proposed in 1994~\cite{pao1994learning}. Primarily, ELM is composed of three successive fully connected layers: input, hidden, and output. The input layer is used to present the input data to the network, whereas the hidden  and output layers conduct the feature extraction and data classification, respectively. When the input data is presented to the network, it gets relayed to the hidden layer, where all the relevant and important features are stochastically extracted~\cite{auerbach2014online}. This is done via projecting the input data to high-dimensional space carried out by a large number of hidden neurons~\cite{huang2014insight}. The features extracted by the hidden layer are further relayed to the output layer where the class label associated with the input is identified. A key feature of ELM is that the training is confined only to the output layer synaptic weights, whereas the hidden layer weights are randomly initialized and left unchanged~\cite{huang2006extreme}. This feature speeds up the training in ELM and makes the algorithm attractive for hardware implementations as there is no need for back-propagation.

\begin{figure}[h!tb]
\centerline{\includegraphics [height = 0.35 \textwidth, width = 0.5 \textwidth]{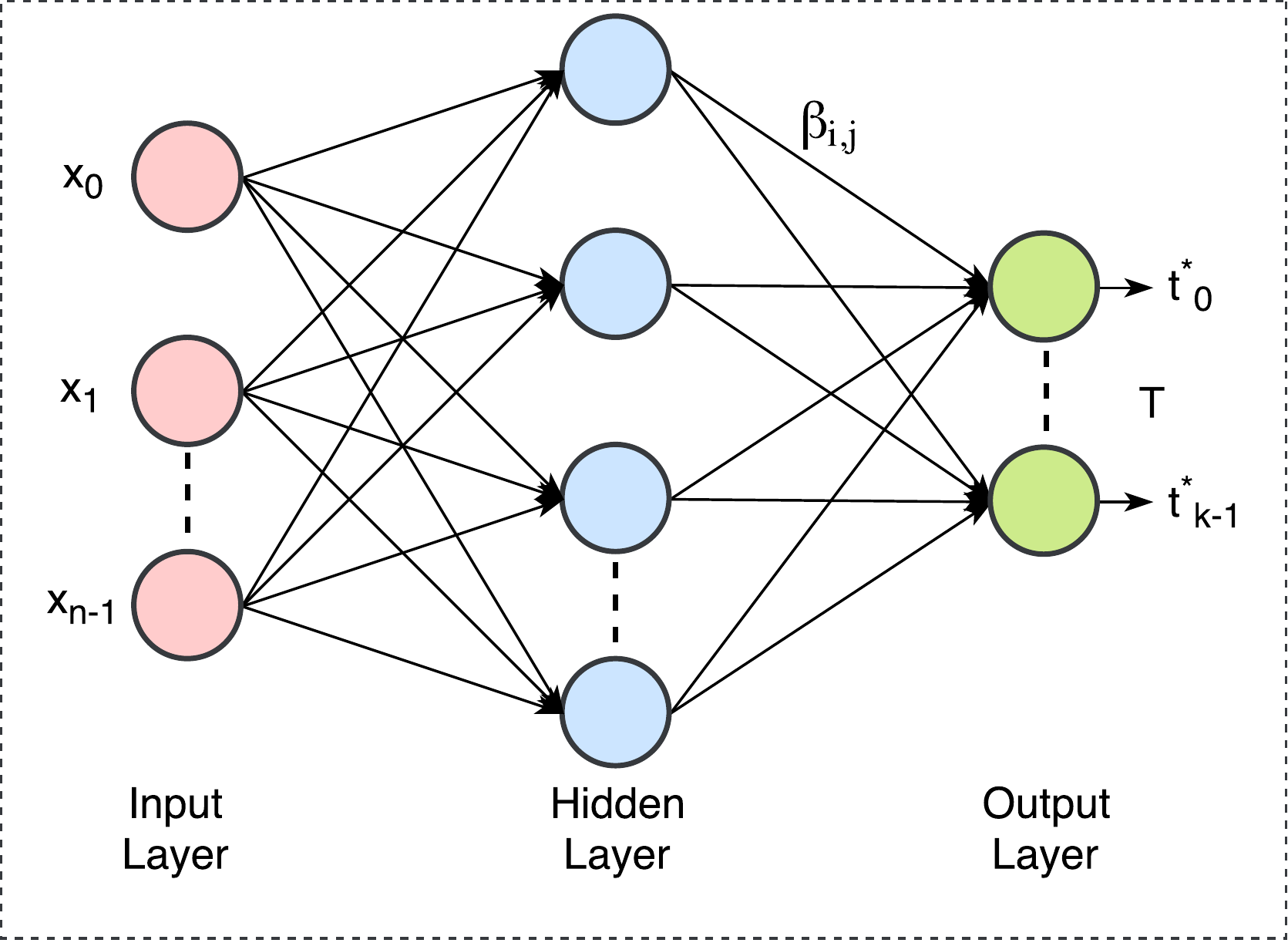}}
\caption{High-level representation of ELM with three layers: an input layer, feature extraction layer (hidden layer), and classification layer (output layer).}
\label{ELM}
\end{figure}

\fig{ELM} illustrates the high-level architecture of an ELM. At runtime, each example in the input dataset is presented to the network as a pair. Each pair contains an input feature vector $X^p$ and its associated class label $t^p$, where $X^p \in \mathbb{R}^{n}$, $\forall p=1, 2, ....,L$ and $L$ is the dataset size. Using~\eq{ff_eqn}, the network feed-forward output can be computed, where $t^*_i$ represents the predicted output of the $i^{th}$ output unit, $\forall i= 1,2, .... k$. $k$ and $\eta$ denote the total number of neurons in the output and hidden layers, $b$ is the bias, while $f$ and $z$ are the activation functions of the hidden and output neurons, respectively. 

\begin{equation}\label{ff_eqn}
    t^*_i = z_{i}\Big(\sum\limits_{j=0}^{\eta-1} \beta_j f_j(X,b)\Big)
\end{equation}

\begin{equation}\label{weight_eqn}
            \beta = H^{-1} T
\end{equation}

Adopting the normal equation, \eq{weight_eqn} (hidden layer output inverse ($H^{-1}$) multiplied by the desired output class labels ($T$)), to find the output layer weight matrix ($\beta$) is a common method in ELM~\cite{huang2004extreme,kasun2013representational} as it offers faster convergence compared to the numerical counterpart. However, realizing the matrix inverse in hardware is cumbersome~\cite{perina2017exploiting}. Rather than using the normal equation, the iterative delta rule algorithm~\cite{jacobs1988increased} is chosen. In delta rule, a weight $\beta_{i,j}$ connecting the $j^{th}$ neuron in hidden layer to $i^{th}$ neuron in output layer is updated according to \eq{weight_eqn2}, where $\alpha$ is the learning rate, $h_{i,j}$ and $(t^{*}_{i} - t^p_{i})$ refer to the input and the output error of the $i^{th}$ neuron, respectively. 

\begin{equation}
\label{weight_eqn2}
    \Delta \beta_{i,j} = \alpha \times h^p_{i,j} \times (t^{*}_{i} - t^p_{i})
\end{equation} 

\section{Design Methodology}
\subsection{Memristive Crossbar Network}
In order to perform the matrix-vector multiplication in ELM, a memristive crossbar is used as it enables high-speed computations while maintaining low power consumption and area overhead. Unfortunately, the memristive crossbar structure offers only positive range of synaptic weights. Therefore, two memristors are used to model the synaptic weights in a bipolar manner. Typically, this is achieved either by using two crossbars or one crossbar with dual input (in this work, dual input refers to a signal and its negation)~\cite{alibart2013pattern,hu2014memristor,hasan2016memristor,chakma2018memristive}. Here, both approaches of realizing the synaptic weights in the memristive crossbar and their constraints will be discussed. Furthermore, in each section, the proposed optimization approach, called semi-trained crossbar, will be investigated.

\subsubsection{Two Crossbars topology} 
\begin{figure}[h!tb]
\centerline{\includegraphics [height = 0.35 \textwidth, width = 0.9 \textwidth]{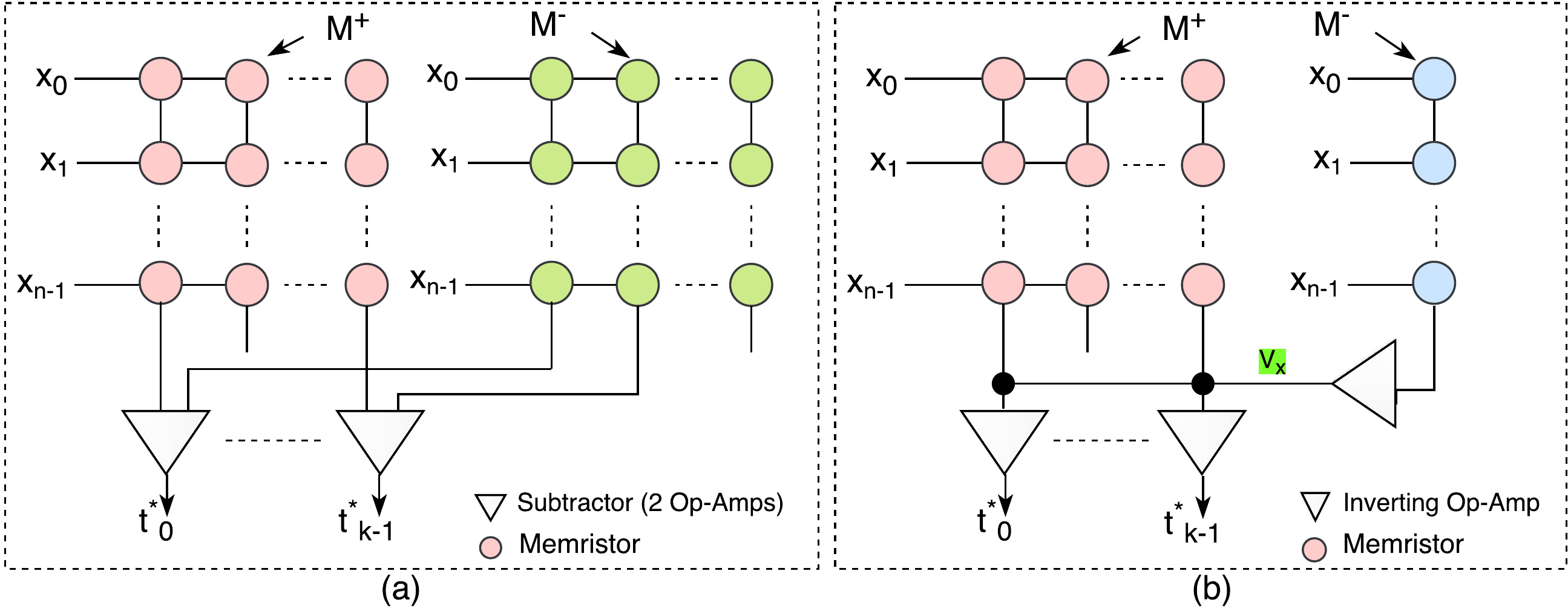}}
\caption{(a) Memristor-based single layer neural network with $k$ number of neurons and $n\times k$ number of synapses modeled as two crossbars. (b) The proposed memristor-based crossbar to model similar single layer network as in (a). The network uses a crossbar to model the synaptic weights, and has an additional untrained column to generate bipolar weights.}
\label{cross_weight}
\end{figure}

In this topology, two crossbars are employed to generate positive and negative weight ranges \cite{alibart2013pattern,hu2014memristor}. \fig{cross_weight}-(a) illustrates the use of two crossbars in emulating the synaptic weights, where each weight value is given by \eq{weight}. $R_{f}$ is the feedback resistance of the Op-Amp based subtracter, and $M^+_{i,j}$ and $M^-_{i,j}$ denote the memristor resistance at the crosspoint ($i,j$) for the left (pink) and right (green) crossbars, respectively. By applying an input voltage ($X^p$ = [$x^p_0, x^p_1, ....x^p_{n-1}$]) at the word-lines (crossbar rows), an output current ($T$ = [$t^*_0, t^*_1,....t^*_{k-1}$]) will be generated as given by \eq{vout}, where $\beta$ is the synaptic weight matrix which can be calculated using \eq{weight}. 

\begin{equation}
\beta_{i,j} = \frac{R_{f}}{M^+_{i,j}} - \frac{R_{f}}{M^-_{i,j}} 
\label{weight}
\end{equation}

\begin{equation}
T = X^p \times \beta
\label{vout}
\end{equation}

It turns out that mapping the synaptic weights to two crossbar arrays overwhelm the learning process, as two crossbars need to be trained rather than one. Moreover, a consistent change in the memristors on the positive and negative crossbars must be sustained to ensure network convergence. Therefore, this research suggests a different approach (called semi-trained) to realize the synaptic weights. This approach utilizes only one crossbar associated with one fixed reference line. The reference line can be created either by using memristors or resistors. \fig{cross_weight}-(b) illustrates the structure of the proposed approach. On the left side, a memristor crossbar is implemented to emulate the synaptic weights. On the right side, one column (denoted by $M^-$) of fixed memristors is used such that positive and negative weight ranges are obtained. By adopting this approach, the number of Op-Amps used at the bit-lines will be shrunk to almost half and thereby reduce hardware resources significantly. 

\subsubsection{One Crossbar topology} 
\begin{figure}[h!tb]
\centerline{\includegraphics [height = 0.35 \textwidth, width = 0.82 \textwidth]{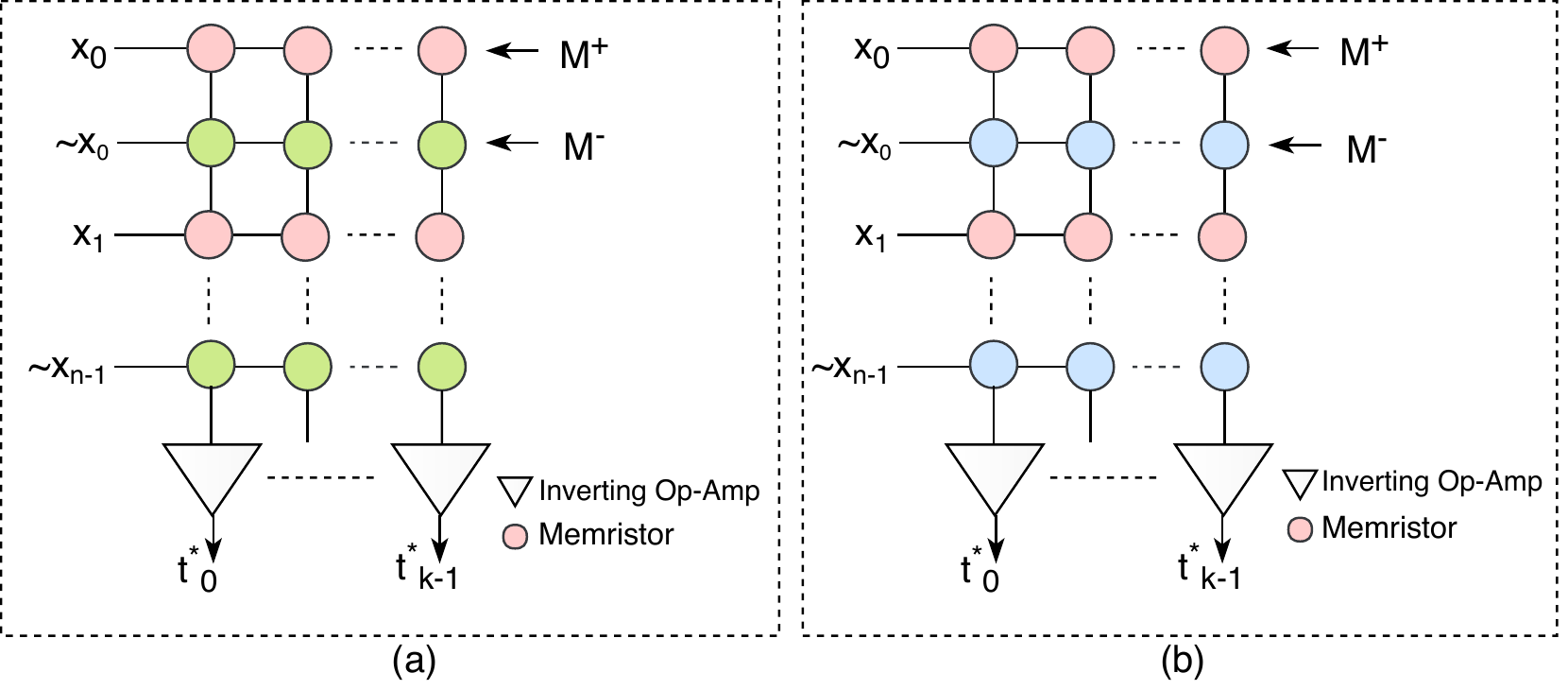}}
\caption{(a) Memristive crossbar to model a single layer neural network with $k$ neurons and $n$ synapses for each neuron. Each synapse ($\beta_{i,j}$) is modeled as two trained memristors ($M_{i,j}^+$ and $M_{i,j}^-$). (b) Proposed memristive crossbar to model a single layer neural network with $k$ neurons and $n$ synapses for each neuron. Two memristors are used to model each synapse, ($\beta_{i,j}$), but one of them is trained ($M_{i,j}^+$) and the other one is fixed ($M_{i,j}^-$).}
\label{cross_weight_appr3}
\end{figure}

One crossbar is used to model the synaptic weights in this topology. However, in this crossbar array, the input needs to be negated to achieve bipolar input-weight matrix-vector multiplication. \fig{cross_weight_appr3} depicts the schematic of one crossbar structure, where the input vector $X^p$ and its negation ($\sim X^p$) are introduced to the crossbar and are multiplied by the synaptic weight matrix $\beta$, as given by \eq{vout}. When it comes to training such a structure, again all the memristors in the crossbar need to be adjusted. By adopting the semi-trained approach, $M^-$ memristors are set to a fixed value, whereas $M^+$ memristors are trained to achieve the desired network convergence. 

\subsubsection{Two Crossbars Vs. One Crossbar:}
\begin{figure}[h!tb]
\centerline{\includegraphics [height = 0.28 \textwidth, width = 0.92 \textwidth]{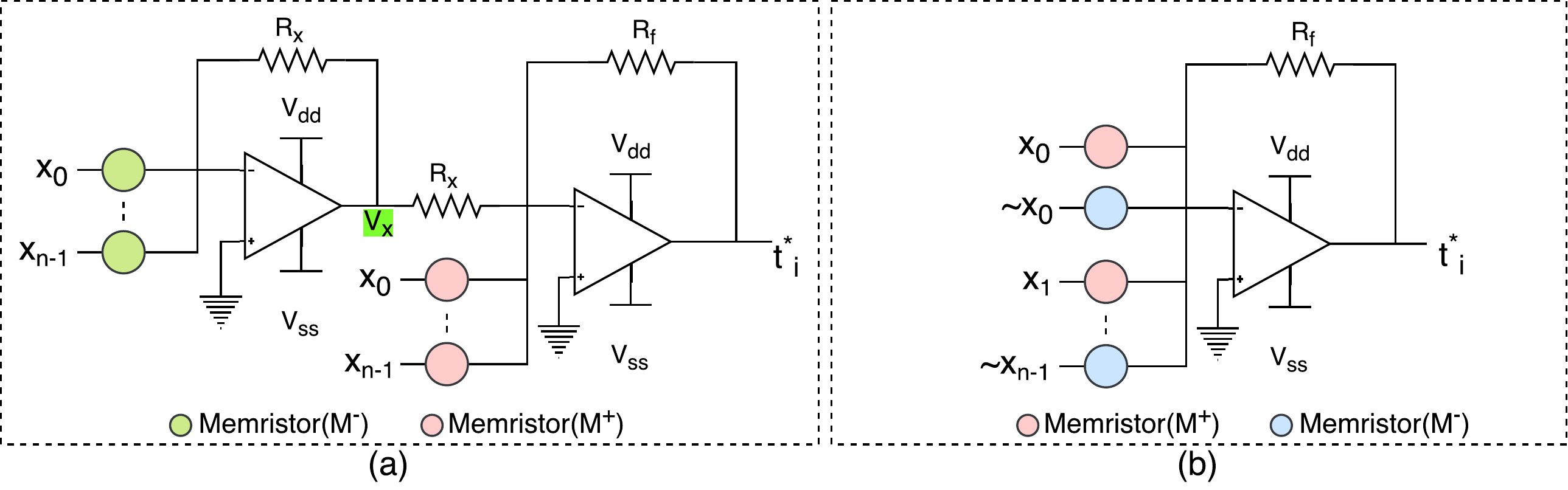}}
\caption{(a) A neuron circuit in a two crossbar topology. (b) A neuron circuit in one crossbar topology.}
\label{cross_weight_down}
\end{figure}

In spite of the fact that both crossbar topologies are capable of performing the intended function (bipolar matrix-vector multiplication), when it comes to hardware, each approach imposes different constraints. The downside of using two crossbars to emulate the synaptic weights is that the input-weight multiplication is fractioned into two parts. One of them is accomplished via $M^+$ crossbar, whereas the second is done in $M^-$ crossbar. Due to this separation, additional constraints are imposed on the network input and its weight range. \fig{cross_weight_down}-(a) illustrates a circuit of one neuron with $n$ inputs, the output of the neuron is given by \eq{neuron_eq} and \eq{neuron_eq2}.   

\begin{equation}
 t^*_i = (x_0 \frac{-R_f}{M^+_0} + ..... +x_{n-1} \frac{-R_f}{M^+_{n-1}}) + V_{x} \frac{-R_f}{R_x}
\label{neuron_eq}
\end{equation}

\begin{equation}
 V_x = x_0 \frac{-R_x}{M^-_0} + ..... +x_{n-1} \frac{-R_x}{M^-_{n-1}}
\label{neuron_eq2}
\end{equation}

Due to the fact that $V_x$ is computed by the first Op-Amp, then its maximum value is always limited to the first Op-Amp biasing voltages ($V_{dd}$ and $V_{ss}$), i.e. \eq{neuron_eq3} must be satisfied. Consequently, the input and weight range will be limited as well as the crossbar size. 

\begin{equation}
 \sum_{i=0}^{n-1} x_i \frac{-R_x}{M^-_i}  \leq (V_{dd} - V_{ss})
\label{neuron_eq3}
\end{equation}

In cases where only one crossbar is used to perform the input-weight matrix-vector multiplication, an additional inverter is needed to negate the input signal. However, in this network the input-weight multiplication will not be segregated. Thus, using only one Op-Amp at the output will suffice, as shown in \fig{cross_weight_down}-b. The output here is given by \eq{neuron_eq_4}. Since $t^*_i$ is associated with one Op-Amp, the constraints that we had in \eq{neuron_eq3} are not applied here. Instead, \eq{neuron_eq_5} must be satisfied, which infers that every single input feature multiplied by its corresponding weight is evaluated separately to be $\leq (V_{dd} - V_{ss})$. This eventually alleviates the constraints we had when using two Op-Amps. Reducing the input and weight range constraints gives more flexibility when it comes to hardware implementation. Moreover, large crossbars can be realized. 

\begin{equation}
 t^*_i = (x_0 \frac{-R_f}{M^+_0} + x_0 \frac{R_f}{M^-_0}..... +x_{n-1} \frac{R_f}{M^-_{n-1}})
\label{neuron_eq_4}
\end{equation}

\begin{equation}
x_i \frac{-R_f}{M^-_i}  \leq (V_{dd} - V_{ss})
\label{neuron_eq_5}
\end{equation}

\begin{figure}[h!tb]
\centerline{\includegraphics [height = 0.5 \textwidth, width = 0.6 \textwidth]{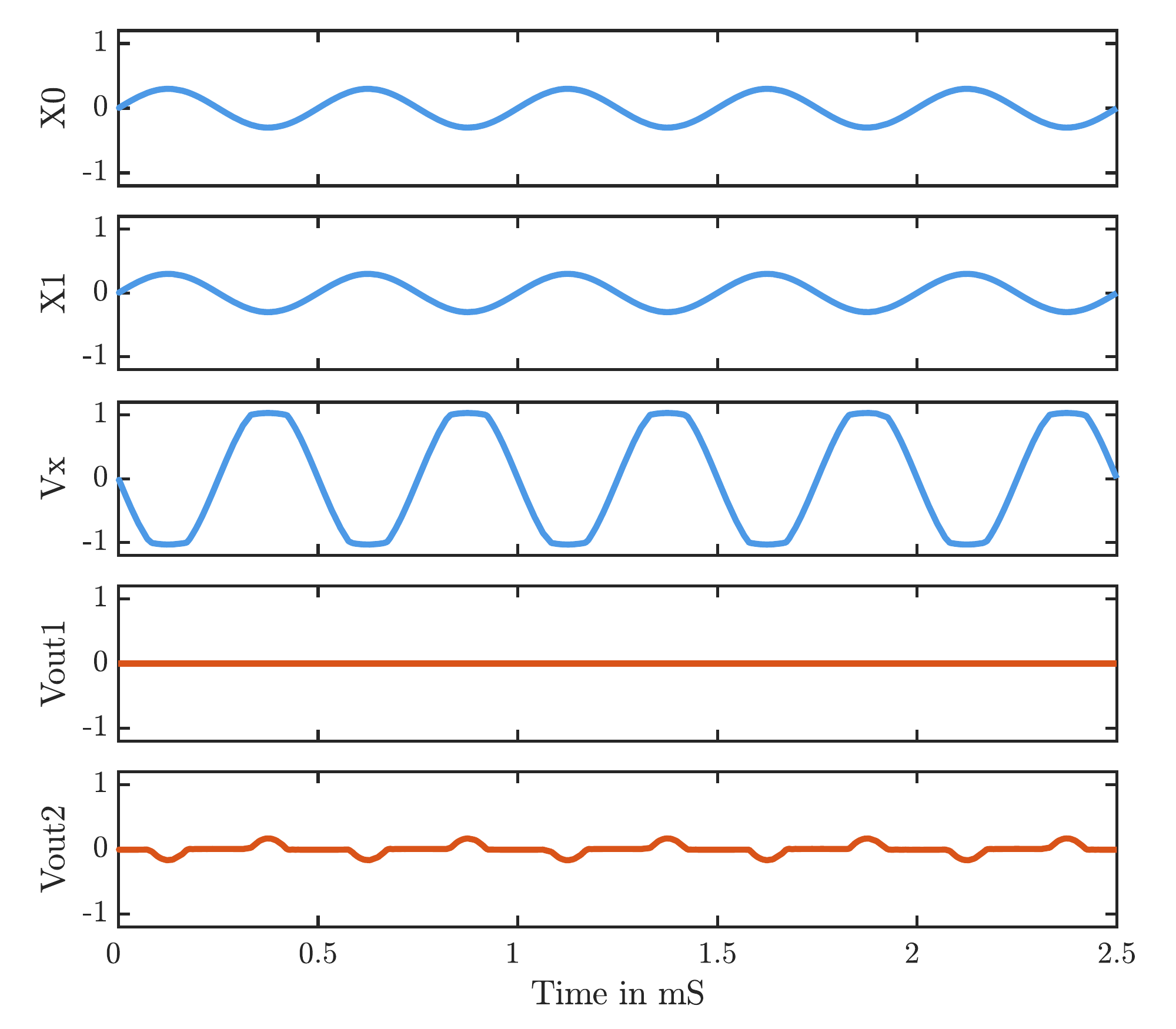}}
\caption{Input-weight multiplication performed using the neuron circuits given in \fig{cross_weight_down}, where all inputs ($X_{n}$) are assigned to $0.3$sin($wt$), $R_f = R_x = 500k\Omega$, and $M^{+} = M^{-} = 250k\Omega$. $V_{out1}$ and $V_{out2}$ denote the output of one and two crossbar neurons, respectively.} 
\label{Neuron}
\end{figure}

\fig{Neuron} demonstrates the input-weight multiplication performed using the neuron circuits from \fig{cross_weight_down}, where all the input features ($X_{n}$, where n =2) are assigned to $0.3$sin($wt$), $R_f = R_x = 500k\Omega$, and $M^{+} = M^{-} = 250k\Omega$. $V_{out2}$ and $V_{out1}$ denote the output of neuron-(a) and neuron-(b), which can be computed based on \eq{neuron_eq} and \eq{neuron_eq_4}, respectively. Although the output in both cases should be the same (=0v), neuron-(a) gives incorrect output as it violated the constraints in \eq{neuron_eq3} (notice that $V_x$ was clipped). This indicates that the neuron circuit in \fig{cross_weight_down}-(b) can handle more input-weight range compared to the former.

\subsection{Delta Rule Algorithm}
In order to simplify the learning process on-chip, delta rule algorithm, described by \eq{weight_eqn2}, is used. However, realizing the delta rule equation in hardware still requires non-trivial resources (subtractor and multiplier). Therefore, this work adopts the simplified delta rule equation used in \cite{hu2014memristor} and shown in \eq{delta_sign}. 

\begin{equation}
 \Delta \beta_{i,j} = \alpha \times S(h^p_{i,j}) \times S(t^{*}_{i} - t^p_{i})
\label{delta_sign}
\end{equation}

\[ S(i) := \left\{
  \begin{array}{l l}
    1 & \quad \text{$i > 0$} \hspace{2mm}\\
    -1 & \quad \text{Otherwise}
  
  \label{permanence_equation2}
\end{array} \right.\]

Here, the weight will essentially change according to the sign of the gradient and has a fixed learning rate. Although such procedure will slow down the convergence speed, the resources used for learning circuity will be significantly minimized. By applying the delta rule to the semi-trained crossbar structure, the iterative change in weight value ensures network convergence can be computed. Recall that each synaptic weight is emulated by fixed and tuned memristors. By using \eq{weight} and \eq{delta_sign}, the net change in the memristor to achieve the desired weight value can be calculated, as in \eq{delta_weight}.

\begin{equation}
 M_{new} = \frac{M_{old} \times R_f}{R_{f} - \alpha \times S(x^p_{i,j}) \times S(t^{*}_{i} - t_{i,p})}
\label{delta_weight}
\end{equation}

\[ M = \frac{1}{M^+} - \frac{1}{M^-} \]

\section{System Design and Analysis}
\begin{figure}[h!tb]
\centerline{\includegraphics [height = 0.6 \textwidth, width = 0.7 \textwidth]{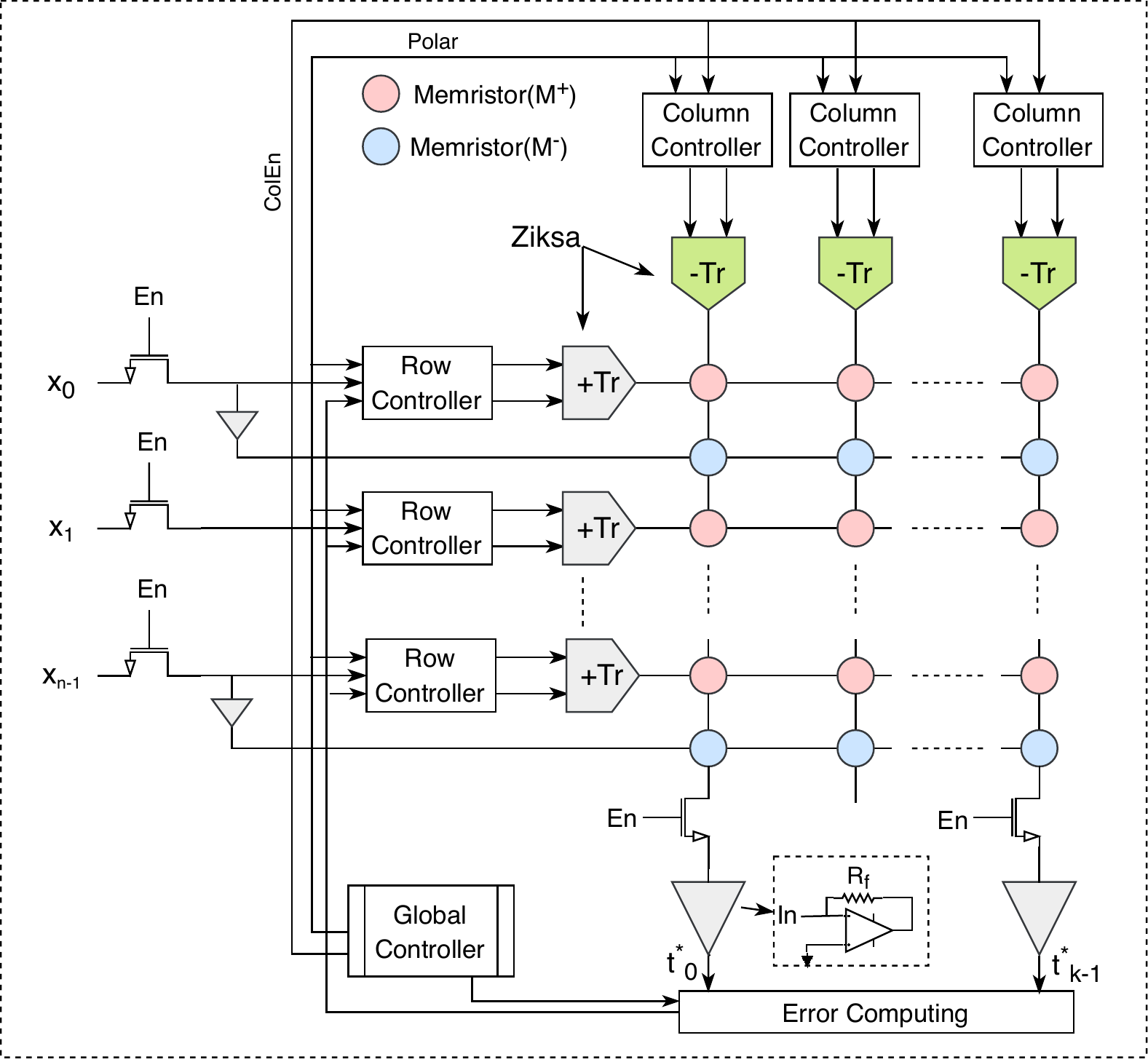}}
\caption{The high-level architecture of the ELM output layer including the memristive crossbar, and the training circuit which is modeled by $+Tr$ and $-Tr$ and controlled by column and row controllers.}
\label{system}
\end{figure}

The system level architecture of ELM consists of two main layers: hidden and output. In this section, the main focus will be on the output layer as it has similar structure to the hidden layer except that the output layer is integrated to the training circuit due to the need of synaptic weight adjustment. \fig{system} shows the architecture of the output layer which essentially has three parts: memristive crossbar, neuron circuit, and training circuit. The memristive crossbar represents a single layer of ELM in which
each column corresponds to one neuron connected
to $R_n$ ($R_n$ is the number of crossbar rows) number of synapses modeled by memristors. The crossbar is responsible for evaluating the input-weight matrix multiplication, whereas the neuron and training circuits are responsible for performing a non-linear transformation on crossbar bit-line outputs and adjusting the weights of the crossbar, respectively.

Primarily, the proposed network runs in two phases: inference and learning. During the inference phase, the input vector ($X^p$ = [$x^p_0, x^p_1, ....x^p_{n-1}$]) is fetched to the network where it gets multiplied by the synaptic weight matrix ($\beta$) to generate the output vector ($T^*$). The output of the network is evaluated by comparing it to input class label and the difference is reported either as logic '1' denoting that $t^*_i > t_i$, or logic '0' otherwise. The output of the error computing unit is stored into a shift register to be used in the learning phase. In the learning phase, the memristor resistances are adjusted according to the sign of the gradient and learning rate. In this work, the tuning of memristor is done column-by-column (training each column takes two clock cycles) through a modified Ziksa training circuit~\cite{zyarah2017ziksa}, which is modeled by $+Tr$ and $-Tr$. Ziksa is used to form an H-Bridge across the memristors that require tuning and by allowing the current to flow through the device in both directions, bipolar weight change can be achieved (more details is provided section~\ref{ziksa_section}). It is important to mention here that each Ziksa unit is controlled by local row and column units which in turn are controlled by a global controller. The global controller determines when to enable the inference and learning phases and is responsible for synchronization. In the following subsections, each unit in the system architecture is discussed in detail.

\subsection{Ziksa: Training Circuitry} \label{ziksa_section}
Recall the simplified delta rule algorithm suggests a fixed adjustment in the memristive weight. In the adopted memristor model, the changes in memristive state variable are dependent on the current flowing in the device as given by~\eq{delta_w}~\cite{kvatinsky2013team}

\begin{equation}
\frac{\Delta w}{\Delta t} = 
\begin{cases}
k_{off}.\Big(\frac{i(t)}{i_{off}} - 1\Big)^{\alpha_{off}}.f_{off}(w),~~~~~0 < i_{off} < i \\
0, ~~~~~~~~~~~~~~~~~~~~~~~~~~~~~~~~~~~~~~~~~~~~~~i_{on} < i < i_{off} \\
k_{on}.\Big(\frac{i(t)}{i_{on}} - 1\Big)^{\alpha_{on}}.f_{on}(w),~~~~~~~~~~~i <i_{on} < 0
\end{cases}
\label{delta_w}
\end{equation}

\noindent where $w$ is the memristor state variable. $k_{off}$, $k_{on}$, $\alpha_{off}$, and $\alpha_{on}$ are constants, $i_{off}$ and $i_{on}$ are the memristor current thresholds, and $f_{on}$ and $f_{off}$ describe the device window function. As the memristor exhibits properties similar to that of a resistor, the current flowing in the device will be limited by its resistance. Thus, to satisfy the learning rule constraints, in this work, we modified our previous design of Ziksa to accommodate this issue. 

\begin{figure}[h!tb]
\centerline{\includegraphics [height = 0.35 \textwidth, width = 0.4 \textwidth]{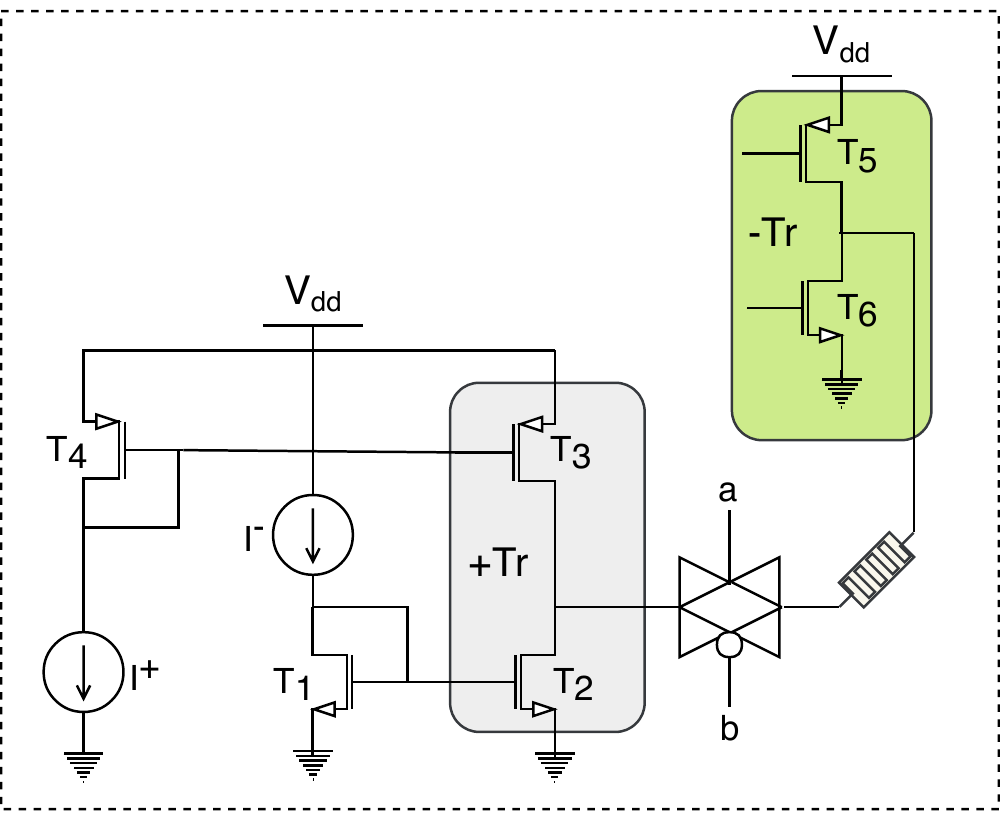}}
\caption{Ziksa unit for adjusting a threshold-current memristor. The unit is mainly composed of four transistors that sandwich the memristor to allow the training current to flow bi-directionally through the memristor. This allows its value to be incremented and decremented.}
\label{Ziksa}
\end{figure}

\fig{Ziksa} illustrates the modified version of Ziksa in which two current mirrors are used to limit the amount of current flowing in the memristor which consequently ensures consistent adjustment of memristor resistance. The circuit works as follows: during the first clock cycle of learning, which involves increasing the weight values (this means decreasing the memristor resistance as $\beta_{i,j} \propto (1 / M_{i,j}$)) in a selected column, current will be provided to the memristor via $T_5$. To ensure a fixed change in the memristor value, this current will be limited to $\approx I^-$ by using a current mirror created by $T_{1-2}$ on the other terminal of the memristor. During the second cycle of training, the weight will be decremented by allowing the current, $\approx I^+$, to flow in the opposite direction.  

\begin{figure}[h!tb]
\centering
\subfigure[]{\includegraphics[width=70mm, height=50mm]{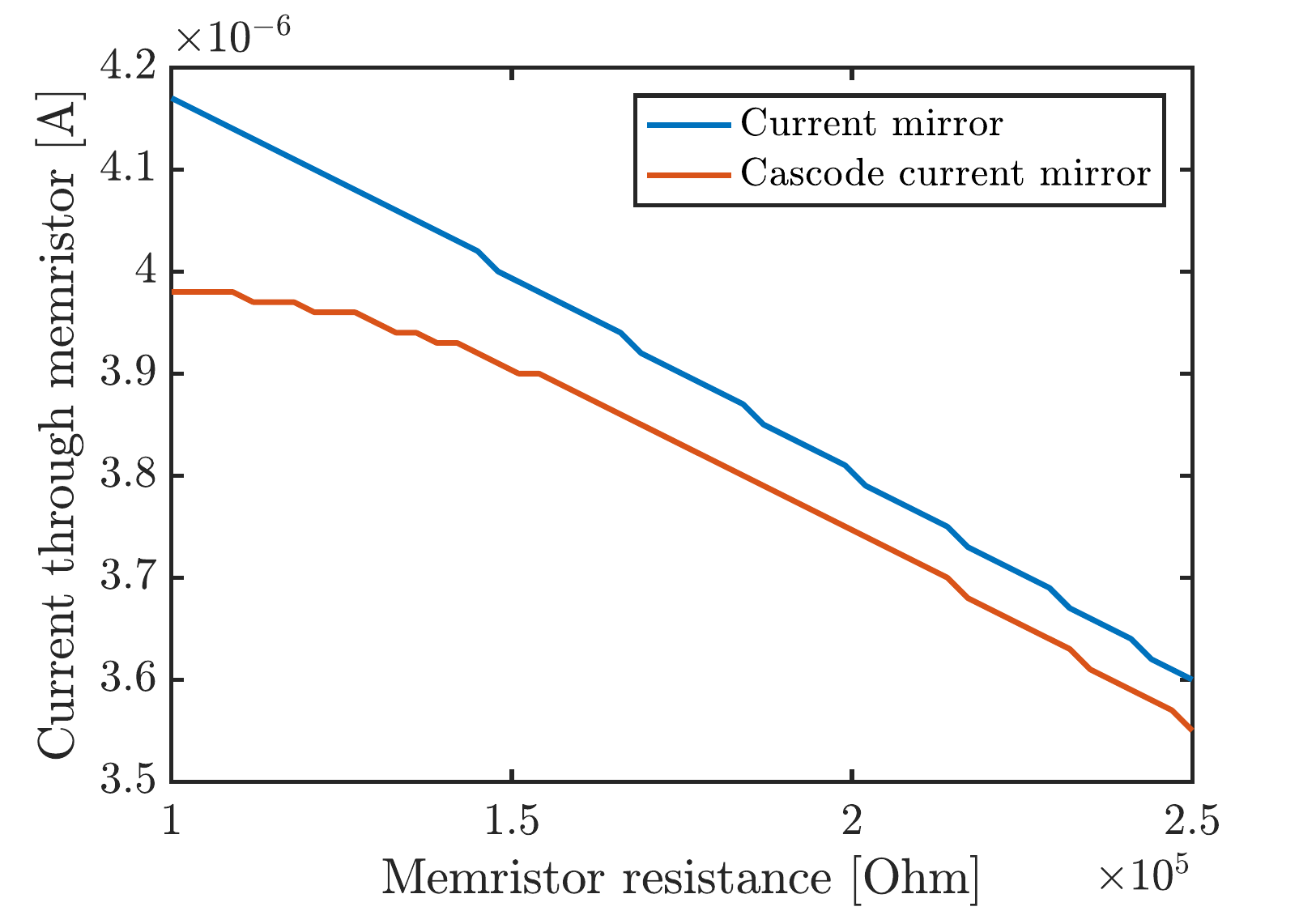}}
\hspace*{-1em}
\subfigure[]{\includegraphics[width=70mm, height=50mm]{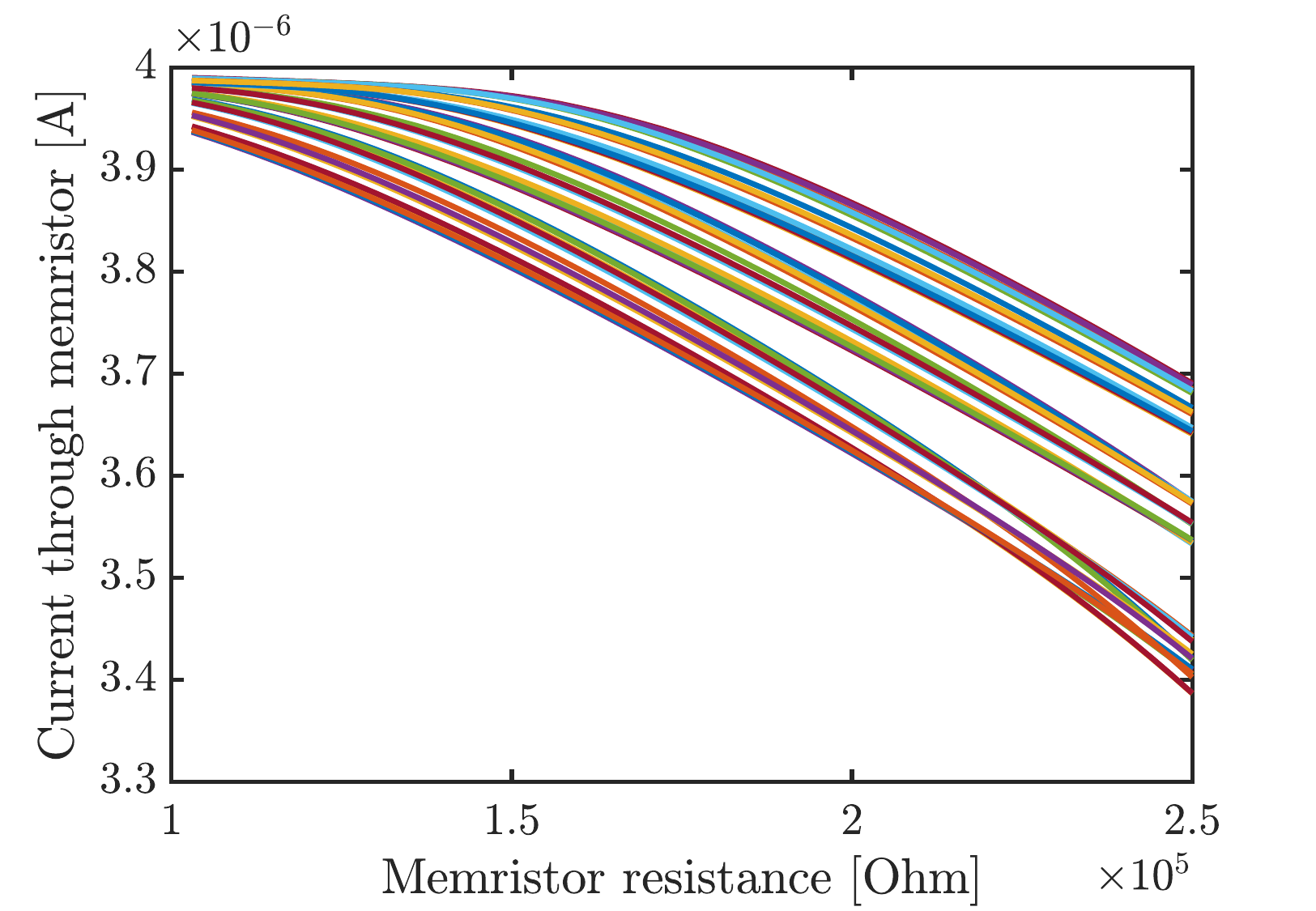}}
\caption{(a) The variation in the current supplied to the memristor while sweeping its value between its low and high resistance states. (b) The variation in the current through the memristor during corner analysis.}
\label{CurrentMirror}
\end{figure}

Practically, fixing the current through the memristor is a difficult condition to meet with the current technology limitations. However, there is still a possibility to limit the variation in the current through memristor while changing its state via a cascode current mirror. The variation in the current through memristor in regular and cascode current mirrors when using $I_{ref}$ = 4 $\upmu$A is depicted in \fig{CurrentMirror}-(a). The corner analysis evaluation while considering the fabrication process, ambient temperature, and supply voltage variations is shown in \fig{CurrentMirror}-(b).

\subsection{Column and Row Local Control Units}
\begin{figure}[h!tb]
\centerline{\includegraphics [height = 0.2 \textwidth, width = 0.6 \textwidth]{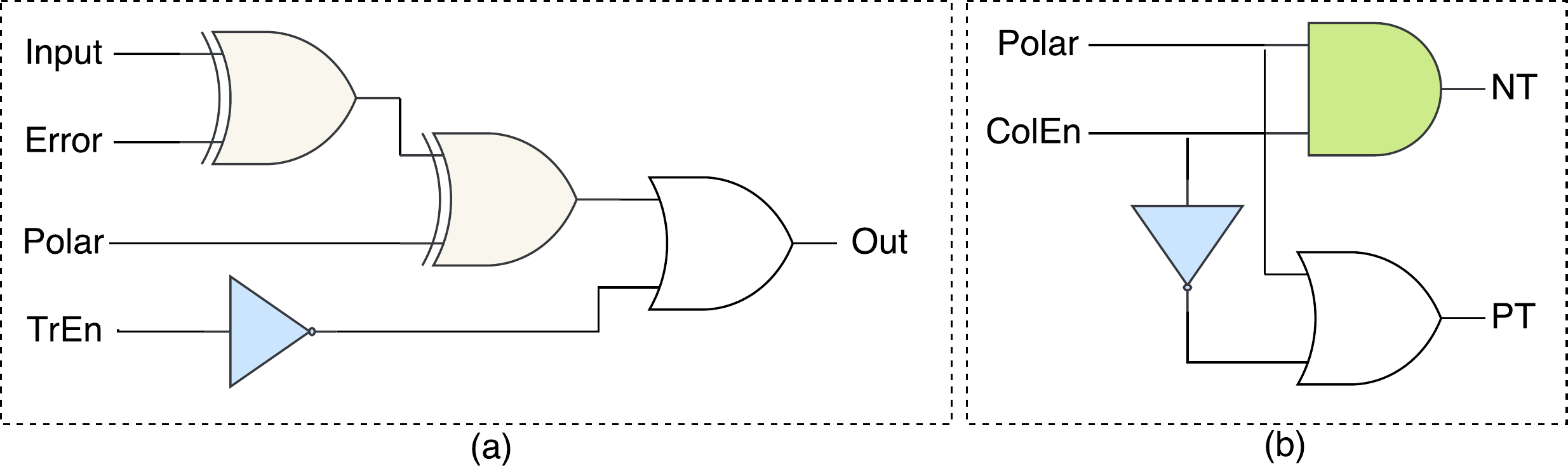}}
\caption{(a) Row control unit to control Ziksa $+Tr$ output. (b) Column control unit to drive Ziksa H-Bridge transistors residing in $-Tr$.}
\label{localCU}
\end{figure}

Ziksa transistors are driven by local control units associated with each column and row. Recall that Ziksa has four transistors that form an H-Bridge sandwiching the memristors in the crossbar. Half of the H-Bridge transistors reside in $+Tr$ and the other half in $-Tr$. Each $-Tr$ is controlled by its associated column controller, shown in \fig{localCU}-(b). The column local control unit consists of a combinational logic circuit that drives the $T_5$ and $T_6$ transistors of $-Tr$. When the learning phase begins, the column controller receives two signals: $ColEn$ and $Polar$. The former signal determines whether a column is selected for training, whereas the latter refers to the system training cycle which can be either positive or negative. During the positive cycle of training, i.e. $Polar$ = '0', all the weights that need to be incremented are adjusted, while the weights required to be decremented are tuned in the negative cycle of training. When a column is enabled by $ColEn$, $T_5$ and $T_6$ transistors of $-Tr$ will be controlled in an alternating way. During the low cycles of the $Polar$ signal, $PT$ is set to low to enable transistor $T_5$, whereas $T_6$ is set to off via $NT$. This allows the current to flow towards the end-terminal of the crossbar row to increase the weight and vice versa for the high period of the $Polar$ signal. In case of $+Tr$, its transistors are controlled via the formed current mirrors with transistors $T_1$ and $T_4$. The output of $+Tr$ is connected to a tri-state gate controlled by the row controller, illustrated in \fig{localCU}-(a). It turns out that the row controller is more complex compared to the column controller because the gradient sign of delta rule is evaluated here via the input signal ($input$) and computed network error ($Error$). Based on the gradient sign, the memristor resistance will be either incremented or decremented. However, this is carried out when the training process is enabled via $TrEn$. According to the $Polar$ signal state, the memristor resistances are modulated during the appropriate training cycle. It is important to mention here that the $ColEn$, $Polar$, and $TrEn$ signals are provided to the local controllers via the global controller.

\subsection{Global Controller}
\begin{figure}[h!tb]
\centerline{\includegraphics [height = 0.4 \textwidth, width = 0.4 \textwidth]{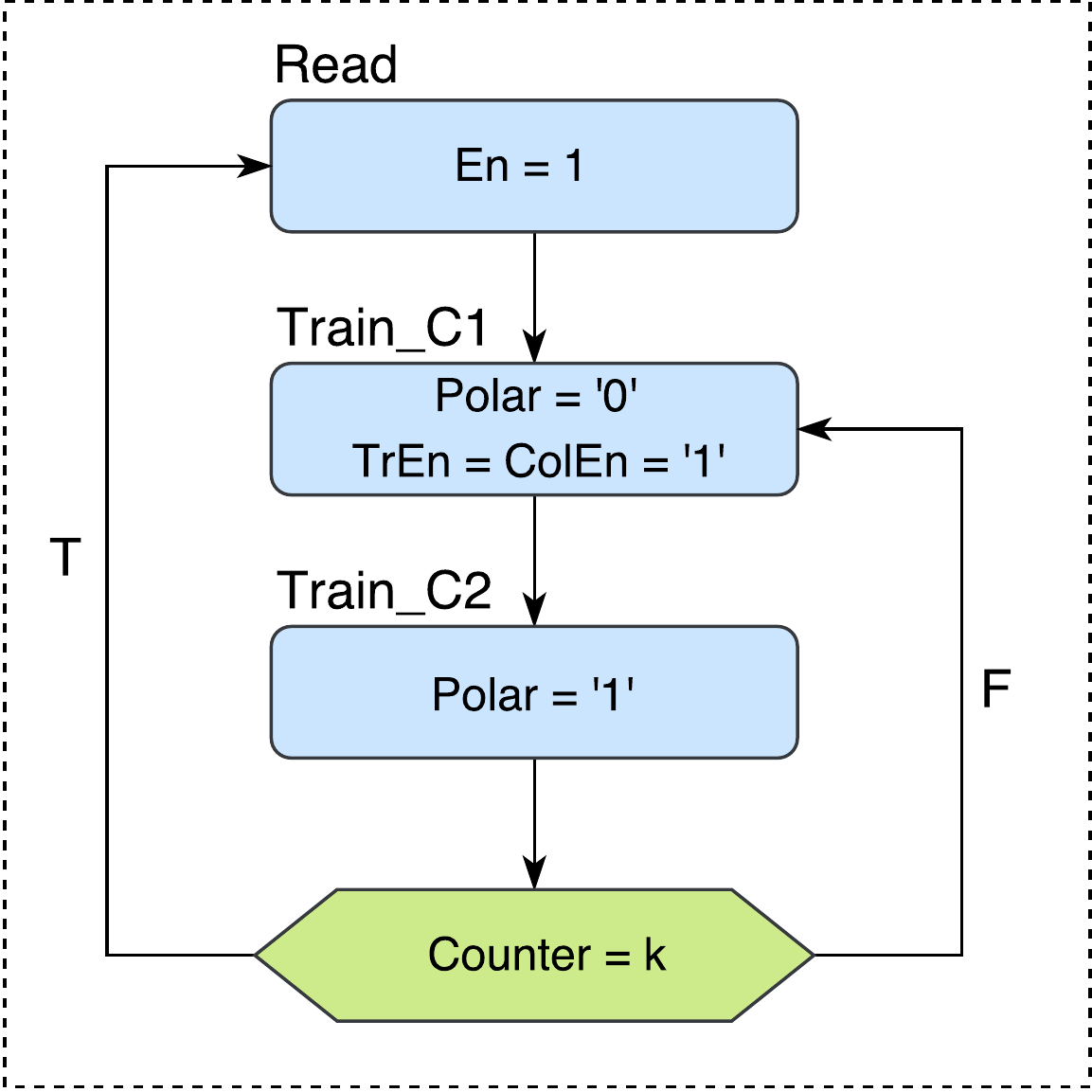}}
\caption{Algorithm state machine flow chart of the global controller.}
\label{Global_ASM}
\end{figure}

All the layers in the proposed system architecture are controlled by a global controller which takes care of data flow and unit synchronization. The global controller runs in three main states. In the first state ($Read$), the global control unit enables the pass transistors to allow the input signals to propagate through the crossbar to perform input-weight matrix multiplication. Once this is done, the output of the network will be captured and the next state ($Train\_C1$) starts. In this state, the first round of training, positive cycle, is performed. Thus, the signals that control the local controllers must be generated. For the column controller, if a column is selected for training, its associated $ColEn$ is set to '1' whereas $Polar$ signal is set to low indicating the positive cycle of training. In case of the row controller, $TrEn$ is set to high, which in association with the input and computed error signs, determines the synaptic weights that need to be incremented. When the global controller runs into the third state ($Train\_C2$), the negative cycle of training will commence, and the same process from the previous cycles will be repeated except that the $Polar$ signal is set to high. The global control unit keeps moving between the second and third states until all the columns of the crossbar are trained. \fig{Global_ASM} is an algorithm state machine chart demonstrating the transition between the states and the output of each one.

\section{Experimental Setup}
The Verilog-A memristor model proposed by ~\cite{kvatinsky2013team} is employed in this work. The memristor value is set based on the device parameters described in~\cite{fan2014design,kawahara20082} such that it meets the network and technology constraints, shown in~\tb{para_table}. The high resistance state (HRS) of the memristor is selected to be 250 k$\Omega$, so that the voltage developed across the memristor does drive the current mirror transistor out of saturation. On the other hand, to keep the power dissipation through the crossbar network as minimum as possible, the low resistance state (LRS) is set to 100 k$\Omega$.  

\begin{table}[h!tb]
\caption{The parameters used for the ELM network simulation.}
\label{para_table}
\begin{center}
\begin{tabular}{|l|c|}
  \hline                       
\rowcolor{Gray}
\textbf{Parameter} & \textbf{Value} \\ \hline
  Memristor tuning current & 4 $\upmu$A  \\ \hline
  Max. memristor current during the inference & 3 $\upmu$A   \\ \hline
  Input voltage range & $<$ $\left|0.5\right|$ v  \\ \hline
  Current threshold & 3.2 $\upmu$A   \\ \hline
  Memristor low resistance state (LRS) & 100 k$\Omega$ \\ \hline
  Memristor high resistance state (HRS) & 250 k$\Omega$ \\ \hline
\end{tabular}
\end{center}
\end{table}

Recall that the proposed system runs in two phases: inference and learning. In the inference phase no change in memristor values should occur. Since a threshold-current memristor is adopted to model the synaptic weights, the current through these devices should never exceed the device threshold to avoid undesired changes in the memristors. Unlike the inference phase, during the learning phase, the current must cross the device threshold-current to adjust the device resistance. According to the experimental setup used in this work, there is a variation in the current through the memristor during the inference phase. However, this variation is estimated to be $\approx$0.6 $\upmu$A. Thus,   by using a current of 4 $\upmu$A during the learning phase and limiting the input current during the inference phase to be 3 $\upmu$A, no overlap between the two phases can occur. 

The following constraints must be fulfilled for the crossbar pass transistors\footnote{These constraints can be overcome by using transmission gates rather than pass transistors.}:
\begin{itemize}
	\item Limit the input voltage such that $V_{DS} << \mid V_{GS} - V_{Th}\mid$. This ensures that the transistor is working in the triode region and an undistorted signal reaches the memristors. 
	\item Assume that $K(V_{GS} - 2V_{Th}) >> G_{mem}(w)$ such that the conductance of the transistor is higher compared to the memristor conductance. Thus, the voltage at the memristor is $\approx$ input voltage~\cite{soudry2015memristor}.
\end{itemize} 

\begin{table}[ht]
\caption{Summary of classification accuracy for binomial and multinomial datasets}
\label{miss}
\centering
\begin{threeparttable}
\begin{tabular}{|c|c|c|c|}
\hline
\rowcolor{Gray} \textbf{DataSet}  & \textbf{ELM, $\eta$ = 1000} & \textbf{VLSI ELM, $\eta$ = 120} & \textbf{This work,} \\
\rowcolor{Gray}   & \cite{huang2012extreme}\tnotex{tnote:robots-r1}       &  \cite{yao2017vlsi}\tnotex{tnote:robots-r2}  & \textbf{$\eta$ = variable}    \\\hline
Diabetes  & 77.95\%    & 77.09\%  & 72.73\% ($\eta$=65) \\ \hline 
Australian Credit   & 87.89\%    & 87.89\%   & 82.16\% ($\eta$=40) \\ \hline 
Iris   & 96.04\%     & -   & 84.66\% ($\eta$=20) \\ \hline 
MNIST & - & - & 93.53\% ($\eta$=180)\tnotex{tnote:robots-r3}\\ \hline
\end{tabular}
\begin{tablenotes}
\item\label{tnote:robots-r1} Software implementation of ELM.
\item\label{tnote:robots-r2} Non-memristive mixed signal implementation of ELM.
\item\label{tnote:robots-r3} All MNIST images are preprocessed with HOG feature descriptor, described in~\cite{zyarah2017extreme}.
    \end{tablenotes}
  \end{threeparttable}
\end{table}

\section{Experimental Results and Network verification}
\subsection{Network Verification}
\begin{figure}[h!tb]
\centering
\subfigure[]{\includegraphics[width=80mm, height=55mm]{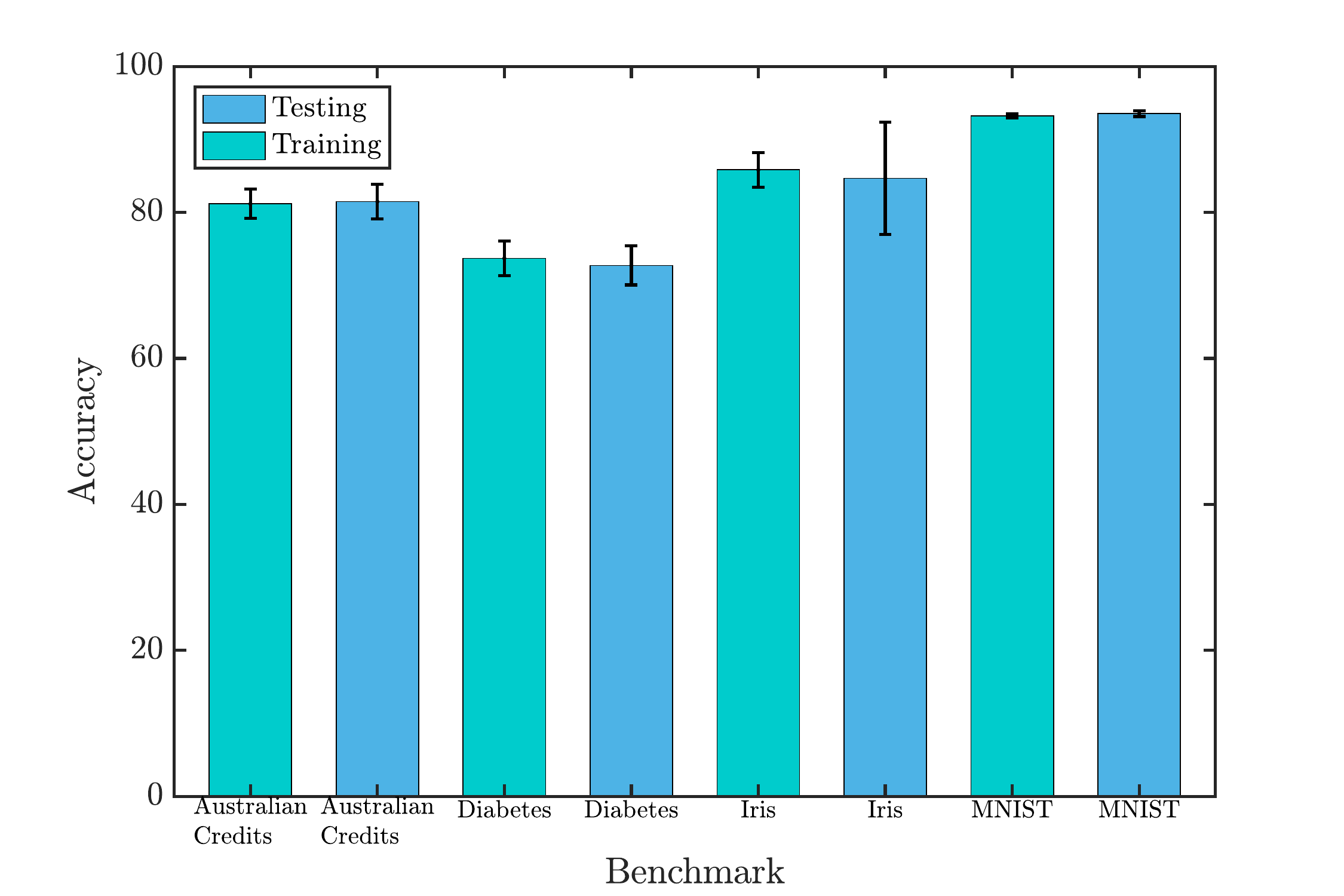}}
\hspace*{-1em}
\subfigure[]{\includegraphics[width=60mm, height=55mm]{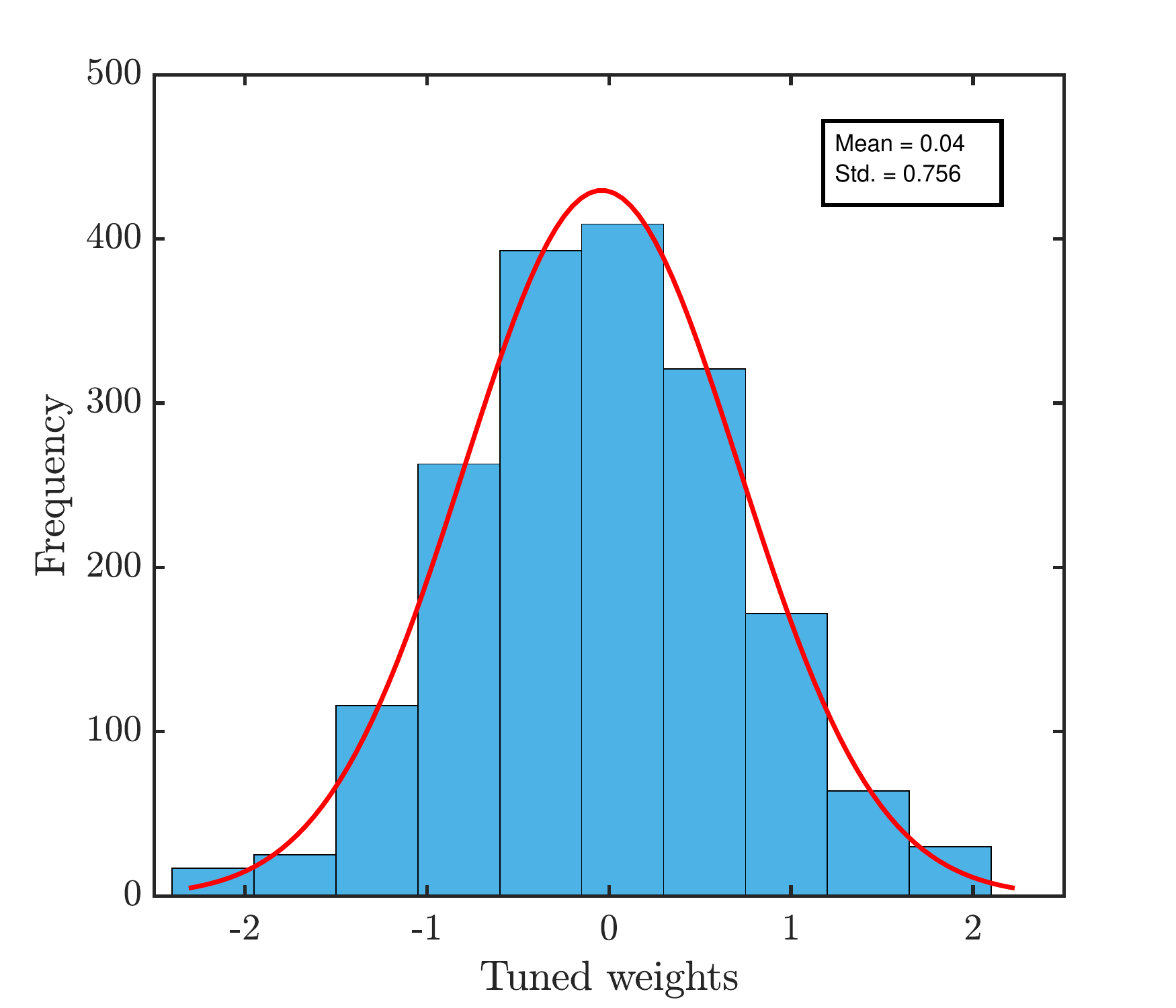}}
\caption{(a) Training and testing accuracy of emulated semi-trained ELM model for binomial and multinomial classifications while considering memristor device variations. (b) Output layer weight distribution of the emulated ELM network trained on MNIST dataset.}
\label{Acc_hist}
\end{figure}

In order to verify the operation of the proposed design, each unit in the network is simulated independently and within a network in Cadence Spectre environment. Then, the same network is emulated in MATLAB and simulated for classification application under the same circuit constraints but with different configurations. The benchmarks employed in this work are selected from UCI library and chosen to be binomial (Diabetes and Australian Credit) and multinomial (Iris). This is added to the multi-class standard hand-written digits dataset, MNIST. \fig{Acc_hist} depicts the weight distribution of the output layer when using MNIST dataset and the achieved accuracy for each dataset during the training and inference phases. Furthermore, the variation in accuracy that may occur due to the process variation of memristors\footnote{10\% variation in memristor resistance range (LRS and HRS) has been considered during the simulation.} and random weight initialization is shown (variation for 5 iterations, each iteration is averaged over 10 runs). \tb{miss} shows a comparison of the achieved accuracy with previous ELM implementations for the same datasets. As can be noticed, although the proposed work offers lower classification accuracy, it has a simpler network as the number of hidden neurons is much lower. However, the performance degradation in the network primarily can be attributed to the input voltage and the weight range constraints imposed on the network. These constraints are due to the memristors limited resistance range and the neuron biasing voltage.

\subsection{Network Scalability}
\begin{figure}[h!tb]
\centerline{\includegraphics [height = 0.32 \textwidth, width = 0.65 \textwidth]{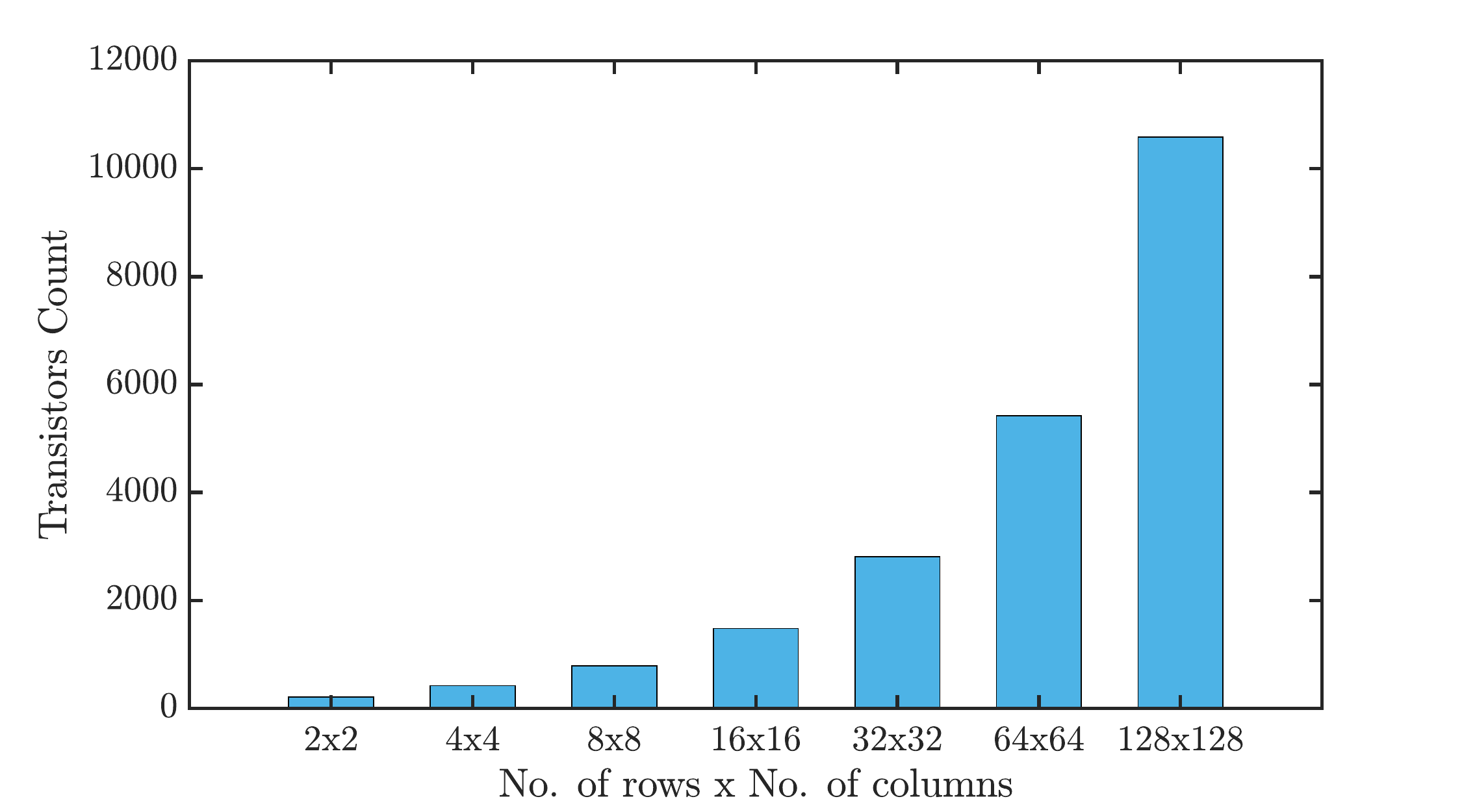}}
\caption{Transistor count for a single layer neural network with various crossbar sizes.}
\label{scale}
\end{figure}

In order to estimate the resources needed to map a large neural network to the proposed design, a full custom design of small-scale (4x4) single layer network is implemented in Cadence using IBM 65nm technology node. \fig{scale} shows the exponential scaling of a single layer neural network from 2x2 up to 128x128, which tends to be linearly proportional to the total transistor count. 

\subsection{Power Consumption}
The total power consumption of the proposed design is estimated for a 4x4 single layer neural network. The power consumption is evaluated in Cadence Spectre environment while running the system at 100 MHz. Considering the worst case scenario, (all crossbar memristors are set to low resistance state), the power consumption is estimated to be 40 $\upmu$W for each 4x4 crossbar and 0.63 $\upmu$W for the digital circuit (the Op-Amp power is not considered). It is important to mention here that the power consumption in a memristor while keeping the device resistance unchanged is assumed to be similar to that of a resistor~\cite{marani2015review}. \tb{power} shows the comparison of power consumption while using different crossbar architectures and different approaches of realizing the synaptic weights. The power consumption of each architecture is achieved by covering all the input combinations and averaging the results. It can be noticed that the semi-trained two-crossbar architecture offers the minimum power consumption as it is compact and uses almost half the number of memristors compared to other designs. 

\begin{table}[ht]
\caption{The power consumption distribution of the proposed crossbar architectures.}
\label{power}
\begin{center}
\begin{tabular}{|l|c|c|c|}
  \hline                       
\rowcolor{Gray} \textbf{Architecture} & \textbf{Digital logic circuit} & \textbf{Crossbar} & \textbf{Total} \\ \hline
  Fully-trained two-crossbar & 1.45 $\upmu$W  & 39.53 $\upmu$W & 40.98 $\upmu$W \\ \hline
  Semi-trained two-crossbar &  1.22 $\upmu$W  & 20.85 $\upmu$W & 22.07 $\upmu$W\\ \hline
  Fully-trained one-crossbar & 1.52 $\upmu$W  & 41.01 $\upmu$W & 42.53 $\upmu$W\\ \hline
  Semi-trained one-crossbar &  1.15 $\upmu$W  & 41.01 $\upmu$W & 42.16 $\upmu$W\\ \hline
\end{tabular}
\end{center}
\end{table}

\section{Conclusions}
This paper investigates a new approach for realizing positive and negative synaptic weights in a crossbar structure using threshold-current memristors. The proposed approach relies on one memristive crossbar to model the weights and uses additional fixed (untrained) columns to generate bipolar weights. Moreover, the paper presents an updated version of the on-device training circuit, Ziksa, which can be used to modulate current-threshold memristors in a crossbar structure. 

The proposed network is tested for classification applications with binomial and multinomial datasets while considering memristor device variations. It is found that the process variations have limited effect on the network performance for large datasets. In these cases, the network has better generalization. In scenarios where power efficiency is a constraint the semi-trained network for two crossbar topology is preferred. Future work will investigate the network performance while considering crossbar resistance, noise effect, and other process variations.

\begin{acks}
The authors would like to thank the members of the Neuromorphic AI research Lab at RIT for their support and critical feedback. The authors would also like to thank the reviewers for their time and extensive feedback to enhance the quality of the paper.
\end{acks}
\input{refer.bbl}


\end{document}

%% file: refer.bbl

%% file: ms.bbl
\begin{thebibliography}{00}


\ifx \showCODEN    \undefined \def \showCODEN     #1{\unskip}     \fi
\ifx \showDOI      \undefined \def \showDOI       #1{{\tt DOI:}\penalty0{#1}\ }
  \fi
\ifx \showISBNx    \undefined \def \showISBNx     #1{\unskip}     \fi
\ifx \showISBNxiii \undefined \def \showISBNxiii  #1{\unskip}     \fi
\ifx \showISSN     \undefined \def \showISSN      #1{\unskip}     \fi
\ifx \showLCCN     \undefined \def \showLCCN      #1{\unskip}     \fi
\ifx \shownote     \undefined \def \shownote      #1{#1}          \fi
\ifx \showarticletitle \undefined \def \showarticletitle #1{#1}   \fi
\ifx \showURL      \undefined \def \showURL       #1{#1}          \fi

\bibitem[\protect\citeauthoryear{Alibart, Zamanidoost, and Strukov}{Alibart
  et~al\mbox{.}}{2013}]%
        {alibart2013pattern}
{Fabien Alibart}, {Elham Zamanidoost}, {and} {Dmitri~B Strukov}. 2013.
\newblock \showarticletitle{Pattern classification by memristive crossbar
  circuits using ex situ and in situ training}.
\newblock {\em Nature communications\/}  {4} (2013).
\newblock


\bibitem[\protect\citeauthoryear{Auerbach, Fernando, and Floreano}{Auerbach
  et~al\mbox{.}}{2014}]%
        {auerbach2014online}
{Joshua~E Auerbach}, {Chrisantha Fernando}, {and} {Dario Floreano}. 2014.
\newblock \showarticletitle{Online extreme evolutionary learning machines}. In
  {\em Artificial Life 14: Proceedings of the Fourteenth International
  Conference on the Synthesis and Simulation of Living Systems}. The MIT Press,
  465--472.
\newblock


\bibitem[\protect\citeauthoryear{Borghetti, Snider, Kuekes, Yang, Stewart, and
  Williams}{Borghetti et~al\mbox{.}}{2010}]%
        {borghetti2010memristive}
{Julien Borghetti}, {Gregory~S Snider}, {Philip~J Kuekes}, {J~Joshua Yang},
  {Duncan~R Stewart}, {and} {R~Stanley Williams}. 2010.
\newblock \showarticletitle{‘Memristive’switches enable ‘stateful’logic
  operations via material implication}.
\newblock {\em Nature\/} {464}, 7290 (2010), 873--876.
\newblock


\bibitem[\protect\citeauthoryear{Chakma, Adnan, Wyer, Weiss, Schuman, and
  Rose}{Chakma et~al\mbox{.}}{2018}]%
        {chakma2018memristive}
{Gangotree Chakma}, {Md~Musabbir Adnan}, {Austin~R Wyer}, {Ryan Weiss},
  {Catherine~D Schuman}, {and} {Garrett~S Rose}. 2018.
\newblock \showarticletitle{Memristive Mixed-Signal Neuromorphic Systems:
  Energy-Efficient Learning at the Circuit-Level}.
\newblock {\em IEEE Journal on Emerging and Selected Topics in Circuits and
  Systems\/} {8}, 1 (2018), 125--136.
\newblock


\bibitem[\protect\citeauthoryear{Chua}{Chua}{1971}]%
        {chua1971memristor}
{Leon Chua}. 1971.
\newblock \showarticletitle{Memristor-the missing circuit element}.
\newblock {\em IEEE Transactions on circuit theory\/} {18}, 5 (1971), 507--519.
\newblock


\bibitem[\protect\citeauthoryear{Fan, Sharad, and Roy}{Fan
  et~al\mbox{.}}{2014}]%
        {fan2014design}
{Deliang Fan}, {Mrigank Sharad}, {and} {Kaushik Roy}. 2014.
\newblock \showarticletitle{Design and synthesis of ultralow energy
  spin-memristor threshold logic}.
\newblock {\em IEEE Transactions on Nanotechnology\/} {13}, 3 (2014), 574--583.
\newblock


\bibitem[\protect\citeauthoryear{Hasan}{Hasan}{2016}]%
        {hasan2016memristor}
{Md~Raqibul Hasan}. 2016.
\newblock {\em Memristor Based Low Power High Throughput Circuits and Systems
  Design}.
\newblock Ph.D. Dissertation. University of Dayton.
\newblock


\bibitem[\protect\citeauthoryear{Hasan, Taha, and Yakopcic}{Hasan
  et~al\mbox{.}}{2017}]%
        {hasan2017chip}
{Raqibul Hasan}, {Tarek~M Taha}, {and} {Chris Yakopcic}. 2017.
\newblock \showarticletitle{On-chip training of memristor crossbar based
  multi-layer neural networks}.
\newblock {\em Microelectronics Journal\/}  {66} (2017), 31--40.
\newblock


\bibitem[\protect\citeauthoryear{Hu, Li, Chen, Wu, Rose, and Linderman}{Hu
  et~al\mbox{.}}{2014}]%
        {hu2014memristor}
{Miao Hu}, {Hai Li}, {Yiran Chen}, {Qing Wu}, {Garrett~S Rose}, {and}
  {Richard~W Linderman}. 2014.
\newblock \showarticletitle{Memristor crossbar-based neuromorphic computing
  system: A case study}.
\newblock {\em IEEE transactions on neural networks and learning systems\/}
  {25}, 10 (2014), 1864--1878.
\newblock


\bibitem[\protect\citeauthoryear{Huang}{Huang}{2014}]%
        {huang2014insight}
{Guang-Bin Huang}. 2014.
\newblock \showarticletitle{An insight into extreme learning machines: random
  neurons, random features and kernels}.
\newblock {\em Cognitive Computation\/} {6}, 3 (2014), 376--390.
\newblock


\bibitem[\protect\citeauthoryear{Huang, Zhou, Ding, and Zhang}{Huang
  et~al\mbox{.}}{2012}]%
        {huang2012extreme}
{Guang-Bin Huang}, {Hongming Zhou}, {Xiaojian Ding}, {and} {Rui Zhang}. 2012.
\newblock \showarticletitle{Extreme learning machine for regression and
  multiclass classification}.
\newblock {\em IEEE Transactions on Systems, Man, and Cybernetics, Part B
  (Cybernetics)\/} {42}, 2 (2012), 513--529.
\newblock


\bibitem[\protect\citeauthoryear{Huang, Zhu, and Siew}{Huang
  et~al\mbox{.}}{2004}]%
        {huang2004extreme}
{Guang-Bin Huang}, {Qin-Yu Zhu}, {and} {Chee-Kheong Siew}. 2004.
\newblock \showarticletitle{Extreme learning machine: a new learning scheme of
  feedforward neural networks}. In {\em Neural Networks, 2004. Proceedings.
  2004 IEEE International Joint Conference on}, Vol.~2. IEEE, 985--990.
\newblock


\bibitem[\protect\citeauthoryear{Huang, Zhu, and Siew}{Huang
  et~al\mbox{.}}{2006}]%
        {huang2006extreme}
{Guang-Bin Huang}, {Qin-Yu Zhu}, {and} {Chee-Kheong Siew}. 2006.
\newblock \showarticletitle{Extreme learning machine: theory and applications}.
\newblock {\em Neurocomputing\/} {70}, 1 (2006), 489--501.
\newblock


\bibitem[\protect\citeauthoryear{Indiveri and Liu}{Indiveri and Liu}{2015}]%
        {indiveri2015memory}
{Giacomo Indiveri} {and} {Shih-Chii Liu}. 2015.
\newblock \showarticletitle{Memory and information processing in neuromorphic
  systems}.
\newblock {\it Proc. IEEE} {103}, 8 (2015), 1379--1397.
\newblock


\bibitem[\protect\citeauthoryear{Jacobs}{Jacobs}{1988}]%
        {jacobs1988increased}
{Robert~A Jacobs}. 1988.
\newblock \showarticletitle{Increased rates of convergence through learning
  rate adaptation}.
\newblock {\em Neural networks\/} {1}, 4 (1988), 295--307.
\newblock


\bibitem[\protect\citeauthoryear{Jo, Chang, Ebong, Bhadviya, Mazumder, and
  Lu}{Jo et~al\mbox{.}}{2010}]%
        {jo2010nanoscale}
{Sung~Hyun Jo}, {Ting Chang}, {Idongesit Ebong}, {Bhavitavya~B Bhadviya},
  {Pinaki Mazumder}, {and} {Wei Lu}. 2010.
\newblock \showarticletitle{Nanoscale memristor device as synapse in
  neuromorphic systems}.
\newblock {\em Nano letters\/} {10}, 4 (2010), 1297--1301.
\newblock


\bibitem[\protect\citeauthoryear{Kasun, Zhou, Huang, and Vong}{Kasun
  et~al\mbox{.}}{2013}]%
        {kasun2013representational}
{Liyanaarachchi Lekamalage~Chamara Kasun}, {Hongming Zhou}, {Guang-Bin Huang},
  {and} {Chi~Man Vong}. 2013.
\newblock \showarticletitle{Representational learning with ELMs for big data}.
\newblock  (2013).
\newblock


\bibitem[\protect\citeauthoryear{Kawahara, Takemura, Miura, Hayakawa, Ikeda,
  Lee, Sasaki, Goto, Ito, Meguro, et~al\mbox{.}}{Kawahara
  et~al\mbox{.}}{2008}]%
        {kawahara20082}
{Takayuki Kawahara}, {Riichiro Takemura}, {Katsuya Miura}, {Jun Hayakawa},
  {Shoji Ikeda}, {Young~Min Lee}, {Ryutaro Sasaki}, {Yasushi Goto}, {Kenchi
  Ito}, {Toshiyasu Meguro}, {and} {others}. 2008.
\newblock \showarticletitle{2 Mb SPRAM (spin-transfer torque RAM) with
  bit-by-bit bi-directional current write and parallelizing-direction current
  read}.
\newblock {\em IEEE Journal of Solid-State Circuits\/} {43}, 1 (2008),
  109--120.
\newblock


\bibitem[\protect\citeauthoryear{Kim, Sah, Yang, Roska, and Chua}{Kim
  et~al\mbox{.}}{2012}]%
        {kim2012neural}
{Hyongsuk Kim}, {Maheshwar~Pd Sah}, {Changju Yang}, {Tam{\'a}s Roska}, {and}
  {Leon~O Chua}. 2012.
\newblock \showarticletitle{Neural synaptic weighting with a pulse-based
  memristor circuit}.
\newblock {\em IEEE Transactions on Circuits and Systems I: Regular Papers\/}
  {59}, 1 (2012), 148--158.
\newblock


\bibitem[\protect\citeauthoryear{Kvatinsky, Friedman, Kolodny, and
  Weiser}{Kvatinsky et~al\mbox{.}}{2013}]%
        {kvatinsky2013team}
{Shahar Kvatinsky}, {Eby~G Friedman}, {Avinoam Kolodny}, {and} {Uri~C Weiser}.
  2013.
\newblock \showarticletitle{TEAM: Threshold adaptive memristor model}.
\newblock {\em IEEE Transactions on Circuits and Systems I: Regular Papers\/}
  {60}, 1 (2013), 211--221.
\newblock


\bibitem[\protect\citeauthoryear{LeCun}{LeCun}{1998}]%
        {lecun1998mnist}
{Yann LeCun}. 1998.
\newblock \showarticletitle{The MNIST database of handwritten digits}.
\newblock {\em http://yann. lecun. com/exdb/mnist/\/} (1998).
\newblock


\bibitem[\protect\citeauthoryear{Lichman}{Lichman}{2013}]%
        {Lichman:2013}
{M. Lichman}. 2013.
\newblock {UCI} Machine Learning Repository.
\newblock   (2013).
\newblock
\showURL{%
\url{http://archive.ics.uci.edu/ml}}


\bibitem[\protect\citeauthoryear{Marani, Gelao, and Perri}{Marani
  et~al\mbox{.}}{2015}]%
        {marani2015review}
{Roberto Marani}, {Gennaro Gelao}, {and} {Anna~Gina Perri}. 2015.
\newblock \showarticletitle{A review on memristor applications}.
\newblock {\em arXiv preprint arXiv:1506.06899\/} (2015).
\newblock


\bibitem[\protect\citeauthoryear{Merkel}{Merkel}{2017}]%
        {merkel2017current}
{Cory Merkel}. 2017.
\newblock \showarticletitle{Current-mode Memristor Crossbars for
  Neuromemristive Systems}.
\newblock {\em arXiv preprint arXiv:1707.05316\/} (2017).
\newblock


\bibitem[\protect\citeauthoryear{Merkel and Kudithipudi}{Merkel and
  Kudithipudi}{2014}]%
        {merkel2014neuromemristive}
{Cory Merkel} {and} {Dhireesha Kudithipudi}. 2014.
\newblock \showarticletitle{Neuromemristive extreme learning machines for
  pattern classification}. In {\em VLSI (ISVLSI), 2014 IEEE Computer Society
  Annual Symposium on}. IEEE, 77--82.
\newblock


\bibitem[\protect\citeauthoryear{Pao, Park, and Sobajic}{Pao
  et~al\mbox{.}}{1994}]%
        {pao1994learning}
{Yoh-Han Pao}, {Gwang-Hoon Park}, {and} {Dejan~J Sobajic}. 1994.
\newblock \showarticletitle{Learning and generalization characteristics of the
  random vector functional-link net}.
\newblock {\em Neurocomputing\/} {6}, 2 (1994), 163--180.
\newblock


\bibitem[\protect\citeauthoryear{Perina, Matias, Marques, Bonato, De~Brito,
  et~al\mbox{.}}{Perina et~al\mbox{.}}{2017}]%
        {perina2017exploiting}
{Andr{\'e}~Bannwart Perina}, {Paulo Matias}, {Eduardo Marques}, {Vanderlei
  Bonato}, {Jo{\~a}o Miguel Gago~Pontes De~Brito}, {and} {others}. 2017.
\newblock \showarticletitle{Exploiting Kant and Kimura’s Matrix Inversion
  Algorithm on FPGA}. In {\em Digital System Design (DSD), 2017 Euromicro
  Conference on}. IEEE, 516--519.
\newblock


\bibitem[\protect\citeauthoryear{Prezioso, Merrikh-Bayat, Hoskins, Adam,
  Likharev, and Strukov}{Prezioso et~al\mbox{.}}{2015}]%
        {prezioso2015training}
{Mirko Prezioso}, {Farnood Merrikh-Bayat}, {BD Hoskins}, {GC Adam},
  {Konstantin~K Likharev}, {and} {Dmitri~B Strukov}. 2015.
\newblock \showarticletitle{Training and operation of an integrated
  neuromorphic network based on metal-oxide memristors}.
\newblock {\em Nature\/} {521}, 7550 (2015), 61--64.
\newblock


\bibitem[\protect\citeauthoryear{Sah, Yang, Kim, and Chua}{Sah
  et~al\mbox{.}}{2012}]%
        {sah2012memristor}
{Maheshwar~Pd Sah}, {Changju Yang}, {Hyongsuk Kim}, {and} {Leon~O Chua}. 2012.
\newblock \showarticletitle{Memristor circuit for artificial synaptic weighting
  of pulse inputs}. In {\em Circuits and Systems (ISCAS), 2012 IEEE
  International Symposium on}. IEEE, 1604--1607.
\newblock


\bibitem[\protect\citeauthoryear{Snider}{Snider}{2008}]%
        {snider2008spike}
{Greg~S Snider}. 2008.
\newblock \showarticletitle{Spike-timing-dependent learning in memristive
  nanodevices}. In {\em Nanoscale Architectures, 2008. NANOARCH 2008. IEEE
  International Symposium on}. IEEE, 85--92.
\newblock


\bibitem[\protect\citeauthoryear{Soudry, Di~Castro, Gal, Kolodny, and
  Kvatinsky}{Soudry et~al\mbox{.}}{2015}]%
        {soudry2015memristor}
{Daniel Soudry}, {Dotan Di~Castro}, {Asaf Gal}, {Avinoam Kolodny}, {and}
  {Shahar Kvatinsky}. 2015.
\newblock \showarticletitle{Memristor-based multilayer neural networks with
  online gradient descent training}.
\newblock {\em IEEE transactions on neural networks and learning systems\/}
  {26}, 10 (2015), 2408--2421.
\newblock


\bibitem[\protect\citeauthoryear{Strukov, Snider, Stewart, and
  Williams}{Strukov et~al\mbox{.}}{2008}]%
        {strukov2008missing}
{Dmitri~B Strukov}, {Gregory~S Snider}, {Duncan~R Stewart}, {and} {R~Stanley
  Williams}. 2008.
\newblock \showarticletitle{The missing memristor found}.
\newblock {\em nature\/} {453}, 7191 (2008), 80--83.
\newblock


\bibitem[\protect\citeauthoryear{Suri, Parmar, Sassine, and Alibart}{Suri
  et~al\mbox{.}}{2015}]%
        {suri2015oxram}
{Manan Suri}, {Vivek Parmar}, {Gilbert Sassine}, {and} {Fabien Alibart}. 2015.
\newblock \showarticletitle{OXRAM based ELM architecture for multi-class
  classification applications}. In {\em Neural Networks (IJCNN), 2015
  International Joint Conference on}. IEEE, 1--8.
\newblock


\bibitem[\protect\citeauthoryear{Taha, Hasan, and Yakopcic}{Taha
  et~al\mbox{.}}{2014}]%
        {taha2014memristor}
{Tarek~M Taha}, {Raqibul Hasan}, {and} {Chris Yakopcic}. 2014.
\newblock \showarticletitle{Memristor crossbar based multicore neuromorphic
  processors}. In {\em System-on-Chip Conference (SOCC), 2014 27th IEEE
  International}. IEEE, 383--389.
\newblock


\bibitem[\protect\citeauthoryear{Yao and Basu}{Yao and Basu}{2017}]%
        {yao2017vlsi}
{Enyi Yao} {and} {Arindam Basu}. 2017.
\newblock \showarticletitle{VLSI extreme learning machine: A design space
  exploration}.
\newblock {\em IEEE Transactions on Very Large Scale Integration (VLSI)
  Systems\/} {25}, 1 (2017), 60--74.
\newblock


\bibitem[\protect\citeauthoryear{Zyarah and Kudithipudi}{Zyarah and
  Kudithipudi}{2017}]%
        {zyarah2017extreme}
{Abdullah~M Zyarah} {and} {Dhireesha Kudithipudi}. 2017.
\newblock \showarticletitle{Extreme learning machine as a generalizable
  classification engine}. In {\em Neural Networks (IJCNN), 2017 International
  Joint Conference on}. IEEE, 3371--3376.
\newblock


\bibitem[\protect\citeauthoryear{Zyarah, Soures, Hays, Jacobs-Gedrim, Agarwal,
  Marinella, and Kudithipudi}{Zyarah et~al\mbox{.}}{2017}]%
        {zyarah2017ziksa}
{Abdullah~M Zyarah}, {Nicholas Soures}, {Lydia Hays}, {Robin~B Jacobs-Gedrim},
  {Sapan Agarwal}, {Matthew Marinella}, {and} {Dhireesha Kudithipudi}. 2017.
\newblock \showarticletitle{Ziksa: On-chip learning accelerator with memristor
  crossbars for multilevel neural networks}. In {\em Circuits and Systems
  (ISCAS), 2017 IEEE International Symposium on}. IEEE, 1--4.
\newblock


\end{thebibliography}
